\documentclass[sigconf]{acmart}

\setcopyright{none}
\renewcommand\footnotetextcopyrightpermission[1]{}

\usepackage[most]{tcolorbox}

\usepackage{times}
\usepackage{latexsym}
\usepackage{listings}
\usepackage{geometry}
\usepackage{xcolor}

\usepackage{listings}
\definecolor{codebg}{RGB}{248,248,248}
\definecolor{codegray}{RGB}{90,90,90}
\definecolor{codepurple}{RGB}{160,32,240}
\definecolor{codeblue}{RGB}{0,92,197}
\definecolor{codegreen}{RGB}{0,128,0}
\definecolor{codered}{RGB}{180,0,0}
\lstdefinestyle{arxivprompt}{
  backgroundcolor=\color{codebg},
  basicstyle=\ttfamily\small,
  keywordstyle=\color{codered}\bfseries,
  commentstyle=\color{codegreen},
  stringstyle=\color{codepurple},
  breaklines=true,
  breakatwhitespace=false,
  frame=single,
  rulecolor=\color{codegray},
  framerule=0.3pt,
  columns=fullflexible,
  keepspaces=true,
  showstringspaces=false,
  tabsize=2,
  captionpos=b,
  morekeywords={System,User,Assistant,Input,Output,Instruction}
}

\usepackage[T1]{fontenc}

\usepackage[utf8]{inputenc}
\usepackage{natbib}

\usepackage{microtype}

\usepackage{graphicx}
\usepackage{hyperref}
\usepackage{footmisc}

\usepackage{amsmath} 
\usepackage{textcomp}
\usepackage{xcolor}
\usepackage{graphicx}
\usepackage{subfigure}
\usepackage{colortbl}
\usepackage{pifont}
\usepackage{multirow}
\usepackage{enumitem}
\usepackage{amsmath}
\usepackage{listings}
\usepackage{xcolor}
\usepackage{framed}
\usepackage{fancyhdr}
\usepackage{url}
\usepackage{pifont}
\usepackage[noend]{algpseudocode}
\usepackage{algorithm}
\usepackage{xspace}
\usepackage{booktabs}
 
\usepackage{multirow}
\usepackage{wrapfig}
\usepackage{amsfonts}

\usepackage{tabularx}
\usepackage{booktabs}
\usepackage{lipsum}

\usepackage{graphicx}
\usepackage{subcaption}

\newtcolorbox{answerbox}{
    colback=gray!10,
    colframe=gray!50,
    boxrule=0.5pt,
    arc=1.5pt,
    left=3pt,
    right=3pt,
    top=3pt,
    bottom=3pt,
    boxsep=1pt,
    before skip=3pt,
    after skip=3pt
}

\AtBeginDocument{%
  }

\setcopyright{acmlicensed} 
\copyrightyear{2018} 
\acmYear{2018} 
\acmDOI{XXXXXXX.XXXXXXX} 
\acmConference[Conference acronym 'XX]{Make sure to enter the correct
  conference title from your rights confirmation email}{June 03--05,
  2018}{Woodstock, NY}  
\acmISBN{978-1-4503-XXXX-X/2018/06}  

\begin{document}

\title{Learning Generalizable Multimodal Representations for Software Vulnerability Detection}


\author{%
Zeming Dong\textsuperscript{1},
Yuejun Guo\textsuperscript{2},
Qiang Hu\textsuperscript{*3},
Yao Zhang\textsuperscript{3},
Maxime Cordy\textsuperscript{1},
Hao Liu\textsuperscript{3}, \\
Mike Papadakis\textsuperscript{1}, and
Yongqiang Lyu\textsuperscript{3}%
}
\affiliation{%
  \institution{%
  \textsuperscript{1}University of Luxembourg,
  \textsuperscript{2}Luxembourg Institute of Science and Technology,
  \textsuperscript{3}Tianjin University\\
  }
  \city{}
  \country{}
}

\renewcommand{\shortauthors}{Dong et al.}


\begin{abstract}
Source code and its accompanying comments are complementary yet naturally aligned modalities — code encodes structural logic while comments capture developer intent. However, existing vulnerability detection methods mostly rely on single-modality code representations, overlooking the complementary semantic information embedded in comments and thus limiting their generalization across complex code structures and logical relationships. To address this, we propose \textsc{MultiVul}, a multimodal contrastive framework that aligns code and comment representations through dual similarity learning and consistency regularization, augmented with diverse code-text pairs to improve robustness. Experiments on widely adopted DiverseVul and Devign datasets across four large language models (LLMs) (i.e., DeepSeek-Coder-6.7B, Qwen2.5-Coder-7B, StarCoder2-7B, and CodeLlama-7B) show that \textsc{MultiVul} achieves up to 27.07\% F1 improvement over prompting-based methods and 13.37\% over code-only \emph{Fine-Tuning}, while maintaining comparable inference efficiency.

\end{abstract}

\ccsdesc[500]{Security and privacy~Software security engineering}

\keywords{Software vulnerability detection, Multimodal learning, Cybersecurity}


\maketitle

\section{Introduction}

\begin{figure}[!h]
    \centering
\includegraphics[width=1.0\columnwidth]{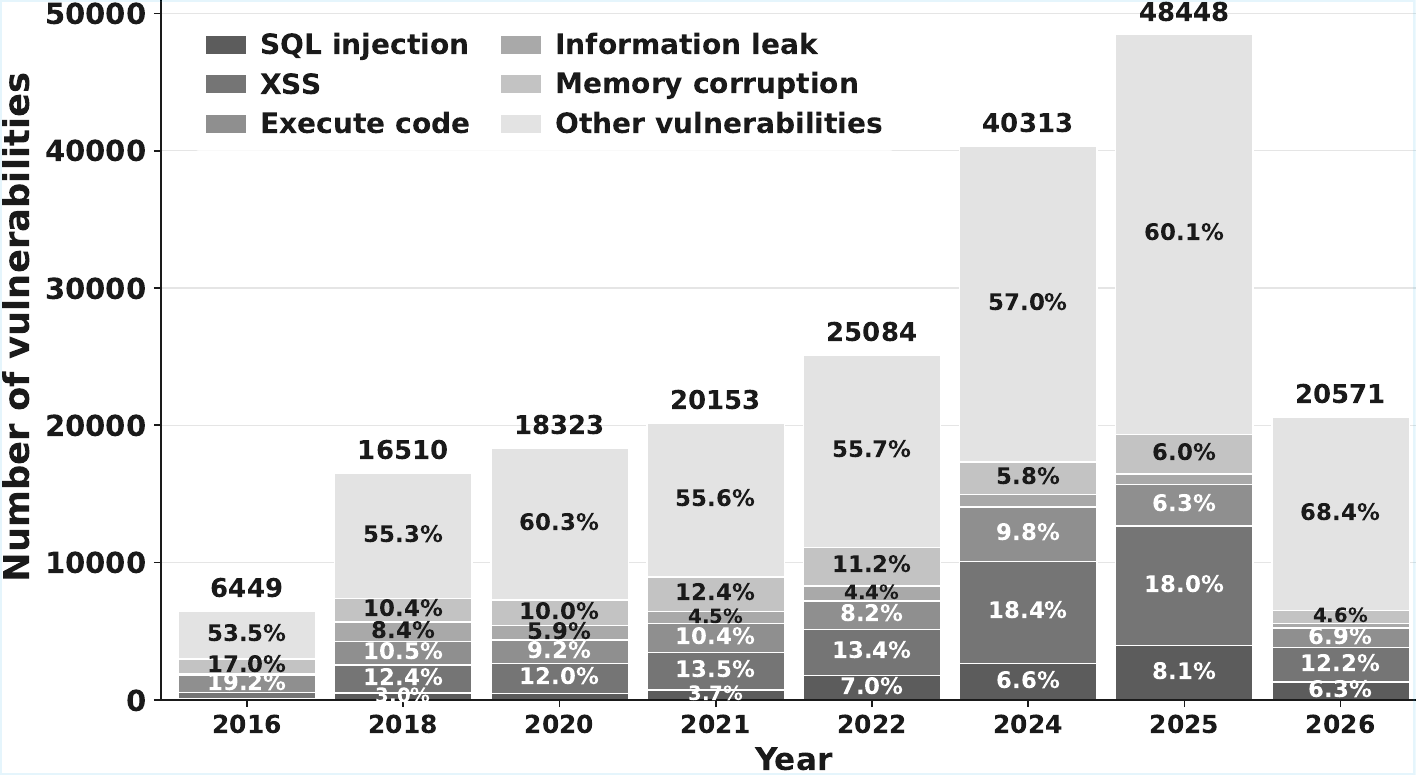}
   \caption{Vulnerabilities by type and year. The statistics are collected from  \emph{CVEdetails}~\cite{cvedetails}.}
   \label{fig:Vul_history}
\end{figure}

Software vulnerability detection is a critical task for ensuring the security and reliability of modern software systems~\cite{chakraborty2021deep}. The need for effective detection has become more urgent as reported vulnerabilities continue to increase in both volume and diversity. As shown in Figure~\ref{fig:Vul_history}, the number of reported vulnerabilities in the first four months of 2026 is already close to the full-year total in 2021, and this growth spans multiple vulnerability types, including memory corruption, SQL injection, and information leakage. This trend indicates that vulnerability detection is becoming increasingly challenging and places growing pressure on existing detection methods. Undetected vulnerabilities may lead to severe consequences, such as unauthorized access and system compromise, causing harm to both users and organizations.

Traditional vulnerability detection methods, including static analysis and dynamic testing, remain indispensable in practice because they provide rule-based and interpretable analyses, and can be highly effective for identifying known vulnerability types in established software security workflows. However, they often fail to capture the complex semantics underlying security issues in source code and face trade-offs in scalability, coverage, and false positives, especially when source code is large, complex, or obfuscated~\cite{sajnani2016sourcerercc,kim2017vuddy,wang2018ccaligner,shiri2024systematic}. To improve automation and scalability, recent work has increasingly turned to learning-based vulnerability detection, leveraging deep neural models (e.g., graph neural networks~\cite{chu2024graph}), pre-trained code models, and large language models (LLMs) to identify vulnerable code types across a wide variety of coding environments, including diverse programming languages and software libraries~\cite{chakraborty2021deep,shiri2024systematic,tao2023vulnerability,zhang2026survey}. Among these, pre-trained code models have become an important foundation for modern code intelligence. Earlier encoder-based models, such as \emph{CodeBERT}~\cite{feng2020codebert} and \emph{GraphCodeBERT}~\cite{guo2020graphcodebert}, established strong baselines for code understanding and downstream software engineering tasks, while recent surveys and empirical studies~\cite{zhang2026survey,sheng2025llms,ding2024vulnerability} suggest that code LLMs (e.g.,  CodeLlama~\cite{roziere2023code}) have emerged as increasingly strong and representative models for a wide range of software engineering tasks. 

Despite this progress, most learning-based vulnerability detection methods still rely primarily on a single code modality, such as token sequences or graph-derived code representations~\cite{chakraborty2021deep,tao2023vulnerability,cheng2022path}. This design still has two important limitations. First, it can restrict detection accuracy, because vulnerability-relevant features are often subtle, distributed across execution paths, and closely tied to program semantics. Sequence-only representations may overlook structural dependencies, while even structure-enhanced representations can remain insufficient for capturing the program behaviors most relevant to vulnerabilities~\cite{guo2020graphcodebert,cheng2022path}. Second, single-modality vulnerability detection methods often generalize poorly. Existing studies show that vulnerability models can learn superficial mappings from source code to labels, overfit to dataset-specific artifacts, and degrade substantially on unseen projects and \emph{out-of-distribution (OOD)} data~\cite {du2024generalization,safdar2025data}. These observations suggest that improving vulnerability detection requires not only stronger code encoders but also richer supervision signals that expose models to higher-level semantics and reduce overfitting to surface features of source code.

Recent studies have explored contrastive and multimodal learning as promising directions for improving semantic modeling in vulnerability detection and related software engineering tasks~\cite{cheng2022path,li2025clever}. These efforts suggest that incorporating information beyond a single code modality can help models learn more informative representations. However, important challenges remain. Existing methods often rely on only a single original code--text view of the input, and lack explicit mechanisms for maintaining robustness under lightweight perturbations of code and text.  Moreover, in some cases, they still depend on natural language descriptions or prompts during inference~\cite{li2025clever}. Consequently, they may struggle to learn code representations that are simultaneously semantically aligned, robust to input variation, and practical for deployment where only code is available. These limitations are particularly important in real-world vulnerability detection, where comments may be missing, noisy, or unavailable at test time, and where models must generalize beyond dataset-specific coding conventions.

To address these limitations, we propose \textsc{MultiVul}, a multimodal vulnerability detection framework that enriches code supervision with automatically generated natural language comments and learns more generalizable code representations through dual-view multimodal training. Specifically, \textsc{MultiVul} combines a dual-encoder architecture, dual-CLIP alignment over original and augmented code--text pairs, and consistency regularization across views. In this way, the model leverages richer semantic supervision during training while retaining a code-only inference pipeline at test time. Different from prior multimodal vulnerability detection methods that rely on textual inputs at inference or align only the single code--text pair, \textsc{MultiVul} addresses a practical and underexplored challenge in vulnerability detection: \textit{how can multimodal supervision be used to improve generalization without assuming the availability of natural language context during deployment?} 

Our key insight is that automatically generated natural language comments can provide complementary semantic information about code functionality and intent, which may be difficult to infer from source code alone. This is particularly useful for vulnerability detection, where security-relevant information can be subtle and depend on API usage patterns, control-flow conditions, and function-level behavior. Prior work has shown that comments can express program behavior in natural language, and that comment augmentation can improve downstream code understanding in code LLMs~\cite{song2024code,cao2025rethinking}. We therefore use generated comments to connect low-level implementation details with higher-level functional behavior, providing richer supervision for vulnerability detection~\cite{rong2025code}. However, simply appending text to code is often insufficient~\cite{xu2023multimodal}. To better use textual supervision, the model should align code and text representations and keep them stable under lightweight perturbations. Existing studies show that lightweight code transformations, such as \emph{GenCode}~\cite{dong2026gencode}, can improve downstream code modeling, while simple textual perturbations such as \emph{random swap} and \emph{random deletion} are also effective data augmentation strategies~\cite{dong2025boosting,wei-zou-2019-eda}. Motivated by these findings, we design a dual-view multimodal training strategy that aligns original and augmented code--text pairs with two CLIP losses and applies consistency regularization to keep nearby views close in the shared embedding space. Together, these mechanisms provide richer training supervision and reduce reliance on shallow code features, thereby improving generalization~\cite{zhang2023generalization,huang2024comparison}.

We evaluate \textsc{MultiVul} on two widely used vulnerability detection benchmarks, DiverseVul~\cite{chen2023diversevul} and Devign~\cite{zhou2019devign}, across four representative code LLMs, including DeepSeek-Coder-6.7B~\cite{guo2024deepseek}, Qwen2.5-Coder-7B~\cite{bai2023qwen}, StarCoder2-7B~\cite{lozhkov2024starcoder}, and CodeLlama-7B~\cite{roziere2023code}. Experimental results show that \textsc{MultiVul} consistently outperforms prompting-based and code-only \emph{Fine-Tuning} baselines, improving F1 by up to 27.07\% and 13.37\%, respectively. Ablations confirm the contribution of augmented alignment and consistency regularization, while OOD and latency results show that \textsc{MultiVul} improves cross-dataset generalization and preserves efficient code-only inference. Our contributions are summarized as follows:
\begin{itemize}[leftmargin=*, itemsep=1pt, topsep=2pt]
\item We propose \textsc{MultiVul}, a multimodal vulnerability detection framework that leverages automatically generated comments during training without requiring text input during inference.
\item We design a dual-view training strategy that aligns original and augmented code--text pairs and stabilizes representations with cross-view consistency regularization.
\item We conduct extensive experiments on two benchmarks and four code LLMs, demonstrating consistent improvements in effectiveness, OOD generalization, and inference efficiency.
\end{itemize}

\section{Background and Related Work}

This section reviews three areas related to \textsc{MultiVul}, including software vulnerability detection, multimodal contrastive learning, and data augmentation for code and text. We highlight how \textsc{MultiVul} differs from existing vulnerability detection methods.

\subsection{Software Vulnerability Detection}
Software vulnerability detection is a long-standing and practically important problem in software security. In real-world software systems, vulnerabilities rarely appear as explicit syntactic errors. Instead, they often arise from semantic errors, implicit developer assumptions, or unsafe interactions among APIs and data manipulations. This makes vulnerability detection particularly challenging in large and evolving projects, where software must interoperate with third-party components, span diverse coding conventions, and satisfy security requirements with incomplete context~\cite {guo2024comprehensive}.

Traditional vulnerability detection methods, such as static analysis and dynamic testing, remain fundamental in security practice because they rely on symbolic execution, formal verification, or fuzzing~\cite{bekrar2011finding}, which provide interpretable analyses and can be highly effective for identifying known types of software weaknesses~\cite{senanayake2023android}. However, they often face well-known limitations in scalability, coverage, and false positives, especially when applied to large and complex source code projects. These challenges have motivated learning-based methods that derive representations directly from source code and use data-driven models for vulnerability detection. Prior work has explored diverse learning-based architectures for vulnerability detection, including task-specific neural models and pre-trained code models. Representative ones include \emph{LineVul}~\cite{fu2022linevul} and \emph{VulBERTa}~\cite{hanif2022vulberta}. More recently, code LLMs (e.g., CodeLlama and StarCoder) have emerged as increasingly strong baselines for vulnerability detection and related software engineering tasks, because they support strong code modeling capacity and adaptability under both fine-tuning and instruction-based prompting~\cite{hou2024large,zhang2026survey,sheng2025llms,ding2024vulnerability}.

At the same time, recent studies have begun to investigate additional modalities (e.g., textual data) beyond a single code modality for improving semantic modeling through contrastive and multimodal learning~\cite{cheng2022path,li2025clever}. These directions are closely related to our work, but differ from it in several important respects.  For example, \emph{CLeVeR}~\cite{li2025clever} relies on vulnerability descriptions during inference, whereas other representative vulnerability detection methods are built on older pre-trained code models or different training formulations, such as CodeBERT-based line-level detection, RoBERTa-based vulnerability pre-training, or multi-task instruction tuning~\cite{li2025clever,fu2022linevul,hanif2022vulberta,du2024generalization}. Directly comparing \textsc{MultiVul} with such methods would make it hard to tell whether performance differences come from the proposed training strategy or simply from differences in inference inputs and base model setups.

Therefore, in the main experiments, we compare \textsc{MultiVul} with strong code-only \emph{fine-tuning} and \emph{prompting-based} baselines using the same code LLM, so that the comparison focuses on the learning framework itself rather than differences in inference input or the underlying code LLM. In addition, we include a standard code–text CLIP baseline in the ablation study to isolate the contribution of basic code–text contrastive alignment relative to \textsc{MultiVul}.

\subsection{Multimodal Contrastive Learning}

Contrastive learning~\cite{hadsell2006dimensionality} has become a widely used paradigm for representation learning in both computer vision (CV) and natural language processing (NLP). As one of the representative methods, CLIP~\cite{radford2021learning} aligns paired image and text representations in a shared embedding space and has shown strong transferability across downstream tasks. Multimodal contrastive learning~\cite{yuan2021multimodal} extends this idea to settings with multiple data modalities. Its core goal is to bring semantically matched inputs from different modalities closer while pushing mismatched pairs apart. In this way, it provides an effective mechanism for aligning representations across modalities, such as image data and text data, and has been shown to improve generalization by encouraging shared semantic structure in the learned embedding space~\cite{zhang2023generalization,huang2024comparison}.

Inspired by these advances, researchers in the software engineering community have recently begun to explore multimodal and contrastive learning for code-related tasks, including vulnerability detection~\cite{ji2024applying,li2025clever}. The main intuition is that additional natural language information (e.g., code comments) can complement source code and help models learn richer semantic representations than code alone. However, applying multimodal contrastive learning to vulnerability detection remains nontrivial. Code and text differ substantially in granularity and information density, making precise semantic alignment difficult in this task~\cite{li2025clever}. Moreover, recent multimodal vulnerability detection studies~\cite{xu2023multimodal} suggest that simple fusion strategies may be insufficient to preserve discriminative security signals under practical conditions. These challenges become especially important when comments are noisy or when deployment must remain code-only rather than rely on description-side inputs.

These observations motivate our design of \textsc{MultiVul}. Instead of relying on a single original code–text pair, \textsc{MultiVul} constructs original and augmented views of both code and text, optimizes their alignment with two CLIP losses, and further stabilizes the shared space with consistency regularization. Unlike previous methods that require textual inputs during inference, \textsc{MultiVul} uses automatically generated and critique-refined code comments only during training and operates on code-only inputs at inference.

\subsection{Data Augmentation}

Data augmentation is a widely used technique for improving model generalization by providing diverse yet related input data variants during training. In NLP, data augmentation methods include lightweight token-level perturbations such as synonym replacement, random swap, and random deletion~\cite{feng2021survey,wei-zou-2019-eda}. These transformation methods are useful because they are simple, label-preserving, and easy to apply at scale~\cite{feng2021survey}.

Data augmentation has also been actively studied in software engineering. Prior work has explored a variety of code transformations, including identifier renaming, refactoring, adversarial training, and interpolation-based synthesis~\cite{gao2020fuzz,yefet2020adversarial,allamanis2021self,bui2021self,wang2022bridging}. These strategies are designed to create alternative views of the same source code while preserving as much of the original semantics as possible~\cite{lacerda2020code}. These studies suggest that data augmentation can be useful for code representation learning, but they also show that not all data augmentation strategies are equally effective or equally practical~\cite{dong2025boosting}.

In particular, existing empirical work reports that traditional code refactoring-based data augmentation often yields limited and inconsistent gains for code modeling~\cite{bielik2020adversarial,yu2022data,dong2024effectiveness,dong2025boosting}. Although semantics-preserving code transformations are intuitively appealing, they are often tied to language-specific transformation rules and nontrivial rewriting pipelines~\cite{lacerda2020code}. Moreover, several data augmentation methods are constrained by the task formulation. For example, \emph{MixCode}~\cite{dong2023mixcode} is designed for classification and relies on one-hot labels during interpolation. These dependencies make such methods less flexible across languages, tools, and downstream tasks. More recent studies revisit simpler perturbation-based data augmentation strategies and show that lightweight transformations (e.g., randomly swapping two statements) can still provide useful training signals for code understanding and robustness, even when they introduce only small local changes rather than fully semantics-preserving rewrites~\cite{dong2025boosting,dong2026gencode}. This observation is especially relevant for \textsc{MultiVul}, where the goal is not to generate fully independent new programs, but to construct nearby views that encourage stable representation learning.

Motivated by these findings, our work adopts augmented views for both code and text. We focus on lightweight perturbations, in particular \emph{random swap (RS)} and \emph{random deletion (RD)}, because they are simple, model-agnostic, and do not rely on language-specific refactoring rules or external program analyses. Moreover, they are also well-suited to contrastive and consistency-based training, as they create nearby but non-identical views of the same sample. In our design, this property matters more than strict semantic preservation, since the augmented views are used to regularize representation learning rather than replace the original data. As a result, RS and RD provide a practical way to construct dual views in \textsc{MultiVul}.

\section{Problem Formulation}
\label{sec:problem_formulation}

We study function-level vulnerability detection. Given a source code function $c$, the detector predicts a binary label $y \in \{0,1\}$, where $y=1$ denotes a vulnerable function and $y=0$ denotes a non-vulnerable function. During training, each function can be paired with an automatically generated natural language comment $t$, forming a multimodal training set:
\begin{equation}
\mathcal{D}=\{(c_i,t_i,y_i)\}_{i=1}^{N},
\end{equation}
where $c_i$ is the source code function, $t_i$ is its generated code comment, $y_i$ is the vulnerability label, and $N$ is the dataset size.

At inference time, the detector receives only the source code function $c_i$. It first obtains a code representation and outputs a vulnerability probability:
\begin{equation}
p_i = f(c_i),
\end{equation}
where $p_i \in (0,1)$ denotes the predicted probability that $c_i$ is vulnerable. The final prediction is obtained by thresholding:
\begin{equation}
\hat{y}_i = \mathbb{I}\!\left[p_i > \delta\right],
\end{equation}
where $\mathbb{I}[\cdot]$ is the indicator function and $\delta$ is the decision threshold, set to $0.5$ by default.

This formulation reflects practical secure code review and repository-scale scanning, where source code is available but high-quality textual context may be missing, noisy, or inconsistent. \textsc{MultiVul} therefore uses multimodal supervision only during training, while preserving code-only inference at deployment time. It does not require comments, prompts, commit messages, vulnerability reports, or other external textual artifacts during inference.

\section{\textsc{MultiVul}}
\begin{figure*}[!h]
    \centering
    \includegraphics[width=1.0\textwidth]{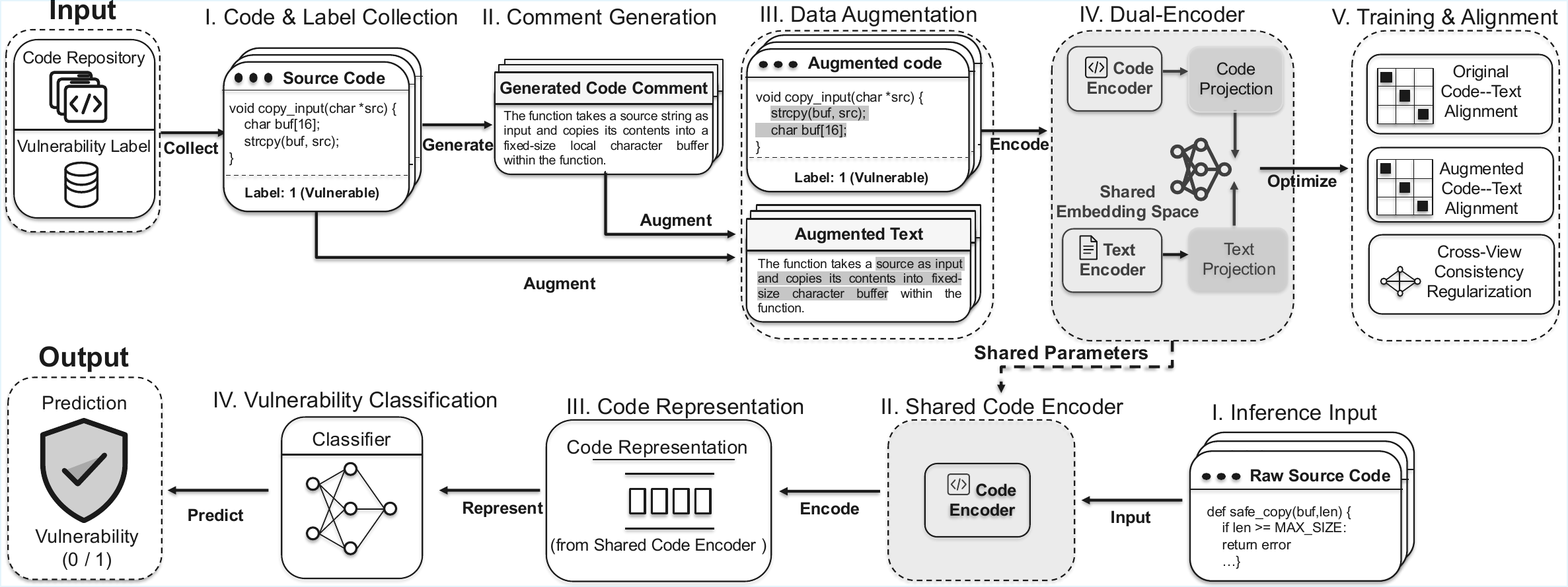}
    \caption{Architecture overview of \textsc{MultiVul}.}
    \label{fig:overview}
\end{figure*}

\textsc{MultiVul} is a multimodal contrastive learning framework for software vulnerability detection. Its core idea is to exploit LLM-generated code comments as auxiliary supervision during training. In this way, the model learns vulnerability-relevant semantics from both code and text, while retaining a lightweight code-only inference pipeline for practical deployment. As illustrated in Figure~\ref{fig:overview}, \textsc{MultiVul} operates in two phases. During training, each source code function and its generated comment are treated as paired multimodal inputs. Both code and text are further augmented to construct an additional view, resulting in original and augmented code--text pairs. These inputs are encoded by a dual-encoder architecture and projected into a shared embedding space. The model is optimized with dual-CLIP alignment, namely original code--text alignment and augmented code--text alignment, together with cross-view consistency that stabilizes learning across different views.

Although the models in \textsc{MultiVul} are decoder-only code LLMs (e.g., Qwen2.5-Coder), we use the term \emph{encoder} to describe their functional role in representation learning, that is, each code LLM maps an input function or comment into a contextual hidden representation, which is then pooled and projected into the shared embedding space. Accordingly, \emph{dual-encoder} in our framework refers to two modality-specific representation producers for code and text, rather than to the original \emph{Transformer} encoder architecture~\cite{vaswani2017attention}. A vulnerability classifier is trained on top of the learned code representation and is applied to code representations during inference.

\subsection{Multimodal Data Construction}
\label{sec:comments_generation}

In real-world software repositories, source code often lacks high-quality natural language comments, which limits the semantic information available to learning-based vulnerability detection~\cite{shi2022evaluation}. To enrich training-time supervision, \textsc{MultiVul} augments each code with an automatically generated comment and further constructs augmented code--text views for multi-view learning.

For each code $c_i$, we apply a strong instruction-tuned code LLM, \emph{Qwen2.5-Coder-32B-Instruct}, to generate a concise one-sentence comment $t_i$ that summarizes the functionality expressed by the code:

\begin{equation}
t_i = \mathrm{LLM\_gen}(c_i).
\end{equation}

To improve faithfulness and reduce speculative descriptions, comment generation follows a critique prompting procedure rather than direct single-pass generation~\cite{sun2024source}. Specifically, the model first produces a draft summary of the code, then reviews its own output to identify unsupported or overly speculative claims, and finally generates a revised one-sentence comment that describes only the functionality explicitly shown in the code. In the default setting used in this work, the prompt explicitly instructs the model not to mention security or vulnerabilities, and not to infer properties such as validation, error handling, permission checks, or safety guarantees unless they are explicitly present in the code. This procedure is fully automatic. More details are provided in Appendix~\ref{sec:appendix_prompting}.

These automatically generated and critique-refined comments provide semantically enriched supervisory information that links code tokens with higher-level program semantics, thereby supporting cross-modal alignment. We additionally generate perturbed code and text views using a combination of \emph{random swap} (RS) and \emph{random deletion} (RD), as lightweight perturbations have been shown in existing empirical studies~\cite{dong2025boosting,dong2026gencode} to provide useful training information for source code representation learning. Compared with more aggressive transformation methods such as back-translation and MixCode~\cite{dong2023mixcode}, RS and RD are simple, model-agnostic perturbations that introduce limited local changes, making them suitable for constructing nearby but non-identical views for contrastive alignment and consistency regularization. Let $\tilde{c}_i$ and $\tilde{t}_i$ denote the augmented code and text, respectively:
\begin{equation}
\tilde{c}_i = \mathrm{Aug}_{\alpha}(c_i), \qquad
\tilde{t}_i = \mathrm{Aug}_{\alpha}(t_i),
\end{equation}
where $\mathrm{Aug}_{\alpha}(\cdot)$ applies RS and RD with augmentation strength $\alpha$. Here, $\alpha$ controls how far the augmented view deviates from the original input.  We study the effect of different $\alpha$ values in Section~\ref{sec:evaluation_sensitivity}.

This stage produces paired original and augmented code--text views for each training instance.  Generated comments provide semantic supervision, while RS and RD  provide controlled view diversity for dual-CLIP alignment and consistency regularization without treating augmented code as an additional supervised sample.
\subsection{Dual-Encoder and Dual-CLIP Alignment}
\label{sec:due-encoders}

Given the training set $\mathcal{D}$ defined in Section~\ref{sec:problem_formulation} and the augmented views $(\tilde{c}_i,\tilde{t}_i)$ introduced in Section~\ref{sec:comments_generation}, \textsc{MultiVul} adopts dual-encoder learning and optimizes the model with dual-CLIP alignment and cross-view consistency regularization. During training, both the original and augmented code--text pairs are used to improve multimodal representation learning.

\paragraph{Dual-encoder learning.}
For each mini-batch of size $B$ that is denoted as
$\{(c_i,t_i,\tilde{c}_i,\tilde{t}_i,y_i)\}_{i=1}^{B}$,
we use a code encoder $f_{\theta}$ and a text encoder $g_{\phi}$ to encode the original and augmented inputs:
\begin{equation}
\begin{aligned}
\mathbf{h}_i^{c} &= f_{\theta}(c_i), &
\mathbf{h}_i^{t} &= g_{\phi}(t_i), \\
\tilde{\mathbf{h}}_i^{c} &= f_{\theta}(\tilde{c}_i), &
\tilde{\mathbf{h}}_i^{t} &= g_{\phi}(\tilde{t}_i),
\end{aligned}
\label{eq:multivul_hidden}
\end{equation}
where $\mathbf{h}_i^{c}$ and $\mathbf{h}_i^{t}$ denote the hidden representations of the original code and text, and
$\tilde{\mathbf{h}}_i^{c}$ and $\tilde{\mathbf{h}}_i^{t}$ denote those of the augmented code and text, respectively.

To perform multimodal contrastive alignment, we project code and text representations into a shared embedding space using modality-specific projection heads, followed by $\ell_2$ normalization:
\begin{equation}
\begin{aligned}
\mathbf{z}_i^{c} &= \mathrm{norm}(W_c \mathbf{h}_i^{c}), &
\mathbf{z}_i^{t} &= \mathrm{norm}(W_t \mathbf{h}_i^{t}), \\
\tilde{\mathbf{z}}_i^{c} &= \mathrm{norm}(W_c \tilde{\mathbf{h}}_i^{c}), &
\tilde{\mathbf{z}}_i^{t} &= \mathrm{norm}(W_t \tilde{\mathbf{h}}_i^{t}),
\end{aligned}
\label{eq:multivul_proj}
\end{equation}
where $W_c$ and $W_t$ are the projection heads shared across views for code and text, respectively.

\paragraph{Dual-CLIP alignment.}
Within each mini-batch of size $B$, \textsc{MultiVul} computes two batch-wise code--text similarity matrices,
$S^{\mathrm{orig}} \in \mathbb{R}^{B\times B}$ and
$S^{\mathrm{aug}} \in \mathbb{R}^{B\times B}$,
for the original and augmented code--text pairs, respectively.
Each entry measures the similarity between the code embedding of the $i$-th sample and the text embedding of the $j$-th sample in the mini-batch:
\begin{equation}
\begin{aligned}
S^{\mathrm{orig}}_{ij} &= \gamma  \, (\mathbf{z}_i^{c})^{\top}\mathbf{z}_j^{t}, \\
S^{\mathrm{aug}}_{ij}  &= \gamma \, (\tilde{\mathbf{z}}_i^{c})^{\top}\tilde{\mathbf{z}}_j^{t},
\end{aligned}
\label{eq:multivul_sim}
\end{equation}
where $i,j \in \{1,\dots,B\}$ index samples within the mini-batch,
$\gamma =\exp(s)$ is a learnable logit scale, and $(\cdot)^{\top}$ denotes the inner product between two $\ell_2$-normalized embeddings.
The diagonal entries correspond to matched code--text pairs from the same training instance, whereas the off-diagonal entries correspond to mismatched pairs within the mini-batch.

Given a similarity matrix $S \in \mathbb{R}^{B\times B}$, we optimize it with a standard symmetric InfoNCE objective~\cite{oord2018representation}:
\begin{equation}
\mathcal{L}_{\mathrm{clip}}(S)
=
\frac{1}{2B}\sum_{i=1}^{B}
\left[
-\log \frac{\exp(S_{ii})}{\sum_{j=1}^{B}\exp(S_{ij})}
-\log \frac{\exp(S_{ii})}{\sum_{j=1}^{B}\exp(S_{ji})}
\right].
\label{eq:multivul_clip}
\end{equation}
The first term performs code-to-text matching by treating the $i$-th code embedding as the query and its paired text embedding as the positive target, while the second term performs the symmetric text-to-code matching.

The dual-CLIP alignment objective is then defined as:
\begin{equation}
\mathcal{L}_{\mathrm{dual\text{-}clip}}
=
\lambda_{\mathrm{clip}}^{\mathrm{orig}}
\mathcal{L}_{\mathrm{clip}}(S^{\mathrm{orig}})
+
\lambda_{\mathrm{clip}}^{\mathrm{aug}}
\mathcal{L}_{\mathrm{clip}}(S^{\mathrm{aug}}),
\label{eq:multivul_dualclip}
\end{equation}
where $\lambda_{\mathrm{clip}}^{\mathrm{orig}}$ and $\lambda_{\mathrm{clip}}^{\mathrm{aug}}$ control the contributions of the original-view and augmented-view alignment losses, respectively.

This dual-CLIP design enforces cross-modal alignment for both the original code--text pair $(c_i,t_i)$ and the augmented pair $(\tilde{c}_i,\tilde{t}_i)$, thereby improving the generalization of the shared embedding space to lightweight perturbations in both code and text.

\paragraph{Cross-view consistency regularization.}
To further stabilize representation learning across the original and augmented views, \textsc{MultiVul} introduces a consistency regularization term in the shared embedding space presented in Figure~\ref{fig:overview}: 
\begin{equation}
\mathcal{L}_{\mathrm{cons}}
=
\frac{1}{2}
\left(
\frac{1}{B}\sum_{i=1}^{B}
\left\lVert
\mathbf{z}_i^{c}-\tilde{\mathbf{z}}_i^{c}
\right\rVert_2^2
+
\frac{1}{B}\sum_{i=1}^{B}
\left\lVert
\mathbf{z}_i^{t}-\tilde{\mathbf{z}}_i^{t}
\right\rVert_2^2
\right),
\label{eq:multivul_cons}
\end{equation}
where $\mathbf{z}_i^{c}$ and $\tilde{\mathbf{z}}_i^{c}$ denote the projected embeddings of the original and augmented code, respectively. $\mathbf{z}_i^{t}$ and $\tilde{\mathbf{z}}_i^{t}$ denote those of the original and augmented text.
Here $\|\cdot\|_2^2$ denotes the squared $\ell_2$ distance, and the factor averages the code-view and text-view consistency terms.
This regularization encourages augmented views to remain close to their original counterparts in the shared embedding space, promoting local smoothness under lightweight perturbations.

\paragraph{Vulnerability Classifier.}
In addition to contrastive alignment, \textsc{MultiVul} trains a supervised binary classifier using the projected code representation.
For each original code input $c_i$, the classifier takes the projected code embedding $\mathbf{z}_i^{c}$ as input and outputs a scalar logit:
\begin{equation}
s_i = \mathrm{MLP}(\mathbf{z}_i^{c}),
\qquad
p_i = \sigma(s_i),
\label{eq:multivul_clslogit}
\end{equation}
where $\mathrm{MLP}(\cdot)$ denotes a multi-layer perceptron classification head, $\sigma(\cdot)$ denotes the sigmoid function, and
$p_i \in (0,1)$ is the predicted probability that $c_i$ is vulnerable.

We optimize the binary cross-entropy loss:
\begin{equation}
\mathcal{L}_{\mathrm{cls}}
=
-\frac{1}{B}\sum_{i=1}^{B}
\left[
y_i \log p_i + (1-y_i)\log(1-p_i)
\right],
\label{eq:multivul_cls}
\end{equation}
where $y_i \in \{0,1\}$ is the ground-truth vulnerability label.

Overall, \textsc{MultiVul} uses dual encoders to learn code and text representations, employs dual-CLIP to align both original and augmented code--text pairs, and applies consistency regularization to stabilize representations across views.



\section{Experimental Setup}
We implement all methods in PyTorch and conduct experiments on a server equipped with 8$\times$ NVIDIA H200 GPUs (143,771 MiB memory per GPU), 2$\times$ Intel Xeon Platinum 8558 CPUs, and 1.8TB RAM. For a fair comparison, all baselines and \textsc{MultiVul} use the same code LLM and data splits. Furthermore, detailed hyperparameters and training configurations are provided in Appendix~\ref{sec:appendix_hyper}.

\textbf{Datasets.} We conduct experiments on two widely used vulnerability detection benchmarks, including \emph{Devign}~\cite{zhou2019devign} and \emph{DiverseVul}~\cite{chen2023diversevul}. Table~\ref{tab:dataset_stats_Appendix} reports the statistics of the datasets used in our experiments across different splits. Specifically, the table includes the number of functions (\#Funcs), average lines of code (\emph{Avg LOC}), average non-empty lines of code (\emph{Avg nLOC}), average number of tokens (\emph{Avg Tokens}), and the number of non-vulnerable and vulnerable samples in each split.

\begin{table*}[!ht]
\centering
\caption{
Dataset statistics used in our experiments. \emph{LOC} counts include blank lines. \emph{nLOC} denotes non-empty lines of code. Ratio is reported as non-vulnerable: vulnerable.
}
\resizebox{0.8\textwidth}{!}{
\begin{tabular}{@{}llrrrrrrr@{}}
\toprule
\textbf{Dataset} &
\textbf{Split} &
\textbf{\#Funcs} &
\textbf{Avg LOC} &
\textbf{Avg nLOC} &
\textbf{Avg Tokens} &
\textbf{\#Non-Vulnerable} &
\textbf{\#Vulnerable} &
\textbf{Ratio} \\
\midrule
\multirow{3}{*}{Devign}
  & Training & 21,854 & 110.56 & 51.01 & 431.11 & 11,886 & 9,968 & 1.19:1 \\
  & Validation   & 2,733  & 110.68 & 51.52 & 433.70 & 1,486  & 1,247 & 1.19:1 \\
  & Test  & 2,731  & 114.18 & 52.65 & 436.96 & 1,485  & 1,246 & 1.19:1 \\
\midrule
\multirow{3}{*}{DiverseVul}
  & Training & 9,984 & 18.60 & 16.43 & 117.26 & 4,992 & 4,992 & 1.00:1 \\
  & Validation   & 1,248 & 18.82 & 16.67 & 118.73 & 624 & 624 & 1.00:1 \\
  & Test  & 1,250 & 18.71 & 16.49 & 117.38 & 625 & 625 & 1.00:1 \\
\bottomrule
\end{tabular}
}
\label{tab:dataset_stats_Appendix}
\end{table*}

\textbf{Models.} We conduct experiments on four representative open-source code LLMs. In particular, we focus on 7B-scale code LLMs in our experiments because they offer a practical balance between effectiveness and deployment cost, which is particularly relevant in real-world industrial scenarios where compute and latency constraints matter. Prior work~\cite{lu2025demystifying} has shown that reducing model size can substantially lower downstream fine-tuning and inference costs, while carefully designed 7B-scale models can still achieve strong performance with efficient inference. 

We adopted CodeLlama-7B~\cite{roziere2023code}, DeepSeek-Coder-6.7B~\cite{guo2024deepseek}, Qwen2.5-Coder-7B~\cite{bai2023qwen}, and StarCoder2-7B~\cite{lozhkov2024starcoder}. These models are all strong code-oriented LLMs, but differ in their pretraining corpora, architectural design, and data coverage. 
CodeLlama is built on the Llama 2 family and is widely used for code generation and understanding. DeepSeek-Coder is trained on a large multilingual code corpus covering many programming languages. Qwen2.5-Coder is a code-specialized continuation-pretrained model over large-scale GitHub repositories, and StarCoder2 is developed by the \emph{BigCode} project and trained on the Software Heritage archive, offering broad coverage of real-world code. 

\textbf{Baselines.} We compare \textsc{MultiVul} with representative \emph{prompting}-based and \emph{training}-based baselines. For prompting-based baselines, we evaluate Zero-shot, One-shot, and Three-shot inference, as well as \emph{Chain-of-Thought (CoT)} prompting variants, including Zero-shot-CoT, One-shot-CoT, and Three-shot-CoT, using the same code LLM. For training-based baselines, we report \emph{Fine-Tuning}, which performs supervised fine-tuning on code inputs only using the same code LLM, without multimodal alignment. All methods are evaluated on the same data splits and with the same metrics. At test time, \textsc{MultiVul} performs code-only inference, as text views are used exclusively during training for multimodal alignment and consistency regularization.

\textbf{Evaluation Metrics.} We formulate software vulnerability detection as a binary classification task, treating \emph{vulnerable} as the positive class. 
Accordingly, we report \textit{Accuracy}, \textit{Precision}, \textit{Recall}, and \textit{F1}, where Precision, Recall, and F1 are computed with respect to the positive class. 
All metrics are computed on the testing dataset using a fixed decision threshold of $0.5$.

\section{Results}
\label{sec:evaluation}
To evaluate the effectiveness and practicality of \textsc{MultiVul}, we conduct comprehensive experiments and aim to answer the following research questions (RQs):

\begin{itemize}[leftmargin=*, itemsep=1pt, topsep=2pt]
    \item \textbf{RQ1:} How effective is \textsc{MultiVul} in detecting software vulnerabilities?
    \item \textbf{RQ2:} How does each key component contribute to \textsc{MultiVul}?
    \item \textbf{RQ3:} How well does \textsc{MultiVul} generalize under cross-dataset distribution shift?
    \item \textbf{RQ4:} How sensitive is \textsc{MultiVul} to the scale of the code comment generation LLM and the data augmentation strength?
    \item \textbf{RQ5:} How efficient is \textsc{MultiVul} during inference?
\end{itemize}

\subsection{RQ1: Effectiveness of \textsc{MultiVul} }
\label{sec:evaluation_effectiveness}

Table~\ref{tab:main_results} summarizes the effectiveness of \textsc{MultiVul} and all baselines on two datasets under four code LLMs.

\begin{table*}[!ht]
\scriptsize
\setlength{\tabcolsep}{1.8pt}
\renewcommand{\arraystretch}{0.8}
\centering
\caption{Comparison of baselines and \textsc{MultiVul} on DiverseVul and Devign across different code LLMs.
All metrics are reported in percentage (\%). \emph{Fine-Tuning} refers to standard supervised fine-tuning on code inputs only.
Bold marks the best value per column within each code LLM. Blue shading highlights the highest Accuracy and F1 for each code LLM–dataset pair.}

\resizebox{\textwidth}{!}{
\begin{tabular}{lcccccccc|cccccccc}
\toprule
\multirow{3}{*}{\textbf{Methods}} &
\multicolumn{8}{c|}{\textbf{DeepSeek-Coder-6.7B}} &
\multicolumn{8}{c}{\textbf{Qwen2.5-Coder-7B}} \\
\cmidrule(lr){2-9}\cmidrule(lr){10-17}
& \multicolumn{4}{c}{\textbf{DiverseVul}} & \multicolumn{4}{c|}{\textbf{Devign}}
& \multicolumn{4}{c}{\textbf{DiverseVul}} & \multicolumn{4}{c}{\textbf{Devign}} \\
\cmidrule(lr){2-5}\cmidrule(lr){6-9}\cmidrule(lr){10-13}\cmidrule(lr){14-17}
& Accuracy & Precision & Recall & F1 & Accuracy & Precision & Recall & F1
& Accuracy & Precision & Recall & F1 & Accuracy & Precision & Recall & F1 \\
\midrule
Zero-shot      & 49.76 & 48.45 & 7.52  & 13.02 & 54.51 & 51.16 & 7.06  & 12.41
               & 50.40 & 63.16 & 1.92  & 3.73  & 54.43 & 55.56 & 0.80  & 1.58 \\
One-shot       & 48.72 & 47.74 & 27.04 & 34.53 & 54.73 & 61.90 & 2.09 & 4.04
               & 50.88 & 52.00 & 22.88 & 31.78 & 54.51 & 57.14 & 1.28 & 2.51 \\
Three-shot     & 49.44 & 47.98 & 13.28 & 20.80 & 55.53 & 56.56 & 11.08  & 18.52
               & 55.12 & 58.21 & 36.32 & 44.73 & 54.51 & 56.25 & 1.44 & 2.82 \\
Zero-shot-CoT  & 48.48 & 49.20 & 93.92 & 64.58 & 51.14 & 45.58 & 26.44 & 33.47
               & 49.92 & 49.96 & 98.72 & 66.34 & 49.52 & 45.45 & 52.97 & 48.93 \\
One-shot-CoT   & 49.68 & 49.84 & 97.92 & 66.06 & 47.69 & 45.82 & \textbf{33.10} & 38.43
               & 49.84 & 49.92 & 98.72 & 66.31 & 50.33 & 45.57 & 45.43 & 45.50 \\
Three-shot-CoT & 50.16 & 50.08 & \textbf{99.52} & 66.63 & 52.75 & 47.72 & 32.92 & 38.96
               & 50.08 & 50.04 & \textbf{99.20} & 66.52 & 50.99 & 45.51 & 37.40 & 41.06 \\
Fine-Tuning    & 78.72 & 81.88 & 73.76 & 77.61 & 56.19 & 55.51 & 20.22 & 29.64
               & 88.72 & 92.61 & 84.16 & 88.18 & 63.07 & 58.51 & 65.65 & 61.88 \\
\cellcolor[HTML]{DBE7FC}\textbf{MultiVul}
               & \cellcolor[HTML]{DBE7FC}\textbf{90.00} & \cellcolor[HTML]{DBE7FC}\textbf{92.04} & \cellcolor[HTML]{DBE7FC}89.92 & \cellcolor[HTML]{DBE7FC}\textbf{90.98}
               & \cellcolor[HTML]{DBE7FC}\textbf{60.95} & \cellcolor[HTML]{DBE7FC}\textbf{65.73} & \cellcolor[HTML]{DBE7FC}30.18 & \cellcolor[HTML]{DBE7FC}\textbf{41.36}
               & \cellcolor[HTML]{DBE7FC}\textbf{93.60} & \cellcolor[HTML]{DBE7FC}\textbf{93.74} & \cellcolor[HTML]{DBE7FC}93.44 & \cellcolor[HTML]{DBE7FC}\textbf{93.59}
               & \cellcolor[HTML]{DBE7FC}\textbf{67.11} & \cellcolor[HTML]{DBE7FC}\textbf{64.75} & \cellcolor[HTML]{DBE7FC}\textbf{76.21} & \cellcolor[HTML]{DBE7FC}\textbf{70.01} \\
\bottomrule
\end{tabular}
}
\resizebox{\textwidth}{!}{
\begin{tabular}{lcccccccc|cccccccc}
\toprule
\multirow{3}{*}{\textbf{Methods}} &
\multicolumn{8}{c|}{\textbf{StarCoder2-7B}} &
\multicolumn{8}{c}{\textbf{ CodeLlama-7B}} \\
\cmidrule(lr){2-9}\cmidrule(lr){10-17}
& \multicolumn{4}{c}{\textbf{DiverseVul}} & \multicolumn{4}{c|}{\textbf{Devign}}
& \multicolumn{4}{c}{\textbf{DiverseVul}} & \multicolumn{4}{c}{\textbf{Devign}} \\
\cmidrule(lr){2-5}\cmidrule(lr){6-9}\cmidrule(lr){10-13}\cmidrule(lr){14-17}
& Accuracy & Precision & Recall & F1 & Accuracy & Precision & Recall & F1
& Accuracy & Precision & Recall & F1 & Accuracy & Precision & Recall & F1 \\
\midrule
Zero-shot      & 47.60 & 48.60 & 83.20 & 61.36 & 50.62 & 38.36 & 13.48 & 19.95
               & 43.76 & 46.31 & 78.24 & 58.18 & 49.96 & 47.12 & 48.65 & 47.87 \\
One-shot       & 47.36 & 48.32 & 76.16 & 59.13 & 51.28 & 44.47 & 27.13 & 33.70
               & 50.16 & 50.08 & 96.96 & 66.05 & 49.82 & 45.37 & 48.80 & 46.02 \\
Three-shot     & 49.44 & 49.64 & 76.96 & 60.35 & 54.73 & 52.14 & 9.79 & 16.49
               & 48.16 & 49.05 & 94.56 & 64.59 & 49.45 & 45.01 & 48.48 & 46.68 \\
Zero-shot-CoT  & 50.16 & 50.09 & \textbf{92.64} & 65.02 & 53.92 & 48.91 & 21.51 & 29.88
               & 48.24 & 49.10 & 95.52 & 64.86 & 53.04 & 47.70 & 30.02 & 36.85 \\
One-shot-CoT   & 53.12 & 51.83 & 88.48 & 65.37 & 48.28 & 45.07 & 60.83 & 51.78
               & 49.36 & 49.68 & 98.72 & 66.10 & 46.45 & 45.44 & \textbf{56.36} & 50.31 \\
Three-shot-CoT & 48.32 & 49.07 & 88.64 & 63.17 & 46.37 & 40.08 & \textbf{65.10} & 49.61
               & 50.08 & 50.04 & \textbf{99.20} & 66.52 & 56.41 & 53.85 & 31.46 & 39.72 \\
Fine-Tuning    & 84.00 & 87.35 & 79.52 & 83.25 & 62.20 & 61.92 & 44.62 & 51.87

               & 87.28 & 89.49 & 84.48 & 86.91 & 60.51 & 59.86 & 40.93 & 48.62 \\
\cellcolor[HTML]{DBE7FC}\textbf{MultiVul}
               & \cellcolor[HTML]{DBE7FC}\textbf{91.84} & \cellcolor[HTML]{DBE7FC}\textbf{93.37} & \cellcolor[HTML]{DBE7FC}91.52 & \cellcolor[HTML]{DBE7FC}\textbf{92.43}
               & \cellcolor[HTML]{DBE7FC}\textbf{64.69} & \cellcolor[HTML]{DBE7FC}\textbf{62.57} & \cellcolor[HTML]{DBE7FC}56.34 & \cellcolor[HTML]{DBE7FC}\textbf{59.29}
               
               & \cellcolor[HTML]{DBE7FC}\textbf{91.68} & \cellcolor[HTML]{DBE7FC}\textbf{93.06} & \cellcolor[HTML]{DBE7FC}90.08 & \cellcolor[HTML]{DBE7FC}\textbf{91.54}
               & \cellcolor[HTML]{DBE7FC}\textbf{64.30} & \cellcolor[HTML]{DBE7FC}\textbf{63.62} & \cellcolor[HTML]{DBE7FC}45.75 & \cellcolor[HTML]{DBE7FC}\textbf{53.22} \\
\bottomrule
\end{tabular}
}
\label{tab:main_results}
\end{table*}

\textbf{Overall analysis.} Across all code LLMs and both datasets, \textsc{MultiVul} consistently achieves the best \emph{Accuracy} and \emph{F1}, demonstrating its effectiveness for software vulnerability detection. Specifically, standard prompting-based methods (e.g., Zero-shot) are unstable and generally underperform, whereas CoT prompting substantially improves \emph{Recall}, often exceeding 90\% on DiverseVul and reaching competitive Recall on Devign (e.g., 65.10\% with Three-shot-CoT on StarCoder2-7B and 56.36\% with One-shot-CoT on CodeLlama-7B). However, these Recall gains usually come at the expense of lower \emph{Precision}, leading to only moderate F1, with CoT prompting-based methods averaging 65.63\% on DiverseVul and 42.04\% on Devign across the four code LLMs. These results suggest that CoT prompting tends to shift predictions toward the vulnerable class, which improves Recall but often reduces Precision. This observation is consistent with previous studies~\cite{nong2024chain,khare2025understanding,mao2025towards} that CoT prompting can strengthen the ability of vulnerability detection by eliciting more explicit intermediate reasoning, while prompt-based LLM detection still exhibits clear limitations in Precision and overall Accuracy.

In contrast, \textsc{MultiVul} substantially outperforms the strongest prompting-based baseline across all code LLMs and datasets. Concretely, on DiverseVul, \textsc{MultiVul} improves F1 by 24.35\%, 27.07\%, 27.06\%, and 25.02\% over the best prompting-based baseline for DeepSeek-Coder-6.7B, Qwen2.5-Coder-7B, StarCoder2-7B, and CodeLlama-7B, respectively. The corresponding Accuracy gains are also substantial, reaching up to 41.52\%. On Devign, \textsc{MultiVul} also consistently improves over the best prompting-based baseline, with the largest F1 and Accuracy gains reaching 21.08\% and 12.60\%, respectively. These results better reflect practical vulnerability detection performance under class imbalance. Although prompting-based methods (e.g., One-shot-CoT) can sometimes achieve high Recall, \textsc{MultiVul} provides a substantially better trade-off between Precision and Recall, leading to stronger overall performance.
We further compare \textsc{MultiVul} with the training-based baseline, namely standard Fine-Tuning. As shown in Table~\ref{tab:main_results}, \textsc{MultiVul} consistently outperforms Fine-Tuning on all four code LLMs and both datasets. For instance, the largest F1 gains over Fine-Tuning reach 13.37\% on DiverseVul and 11.72\% on Devign. Averaged across the four code LLMs, \textsc{MultiVul} improves over Fine-Tuning by 8.15\% and 7.97\% in F1 on DiverseVul and Devign, respectively. The corresponding average gains in Accuracy are 7.10\% and 3.77\%, respectively.

In summary, compared with prompting-based baselines, \textsc{MultiVul} mitigates the high-recall--low-precision behavior commonly observed in CoT-based prompting~\cite{nguyen2023cof}. Relative to supervised fine-tuning on code inputs only, it consistently improves both Accuracy and F1, demonstrating the value of multimodal alignment for software vulnerability detection. Although Devign yields lower absolute scores than DiverseVul across baselines due to its longer and more project-specific functions, \textsc{MultiVul} still consistently improves over code-only Fine-Tuning under this challenging benchmark.

\begin{figure*}[!ht]
\centering
\subfigure[Zero-shot]{
\begin{minipage}[t]{0.33\linewidth}
\centering
\includegraphics[width=1.0\linewidth]{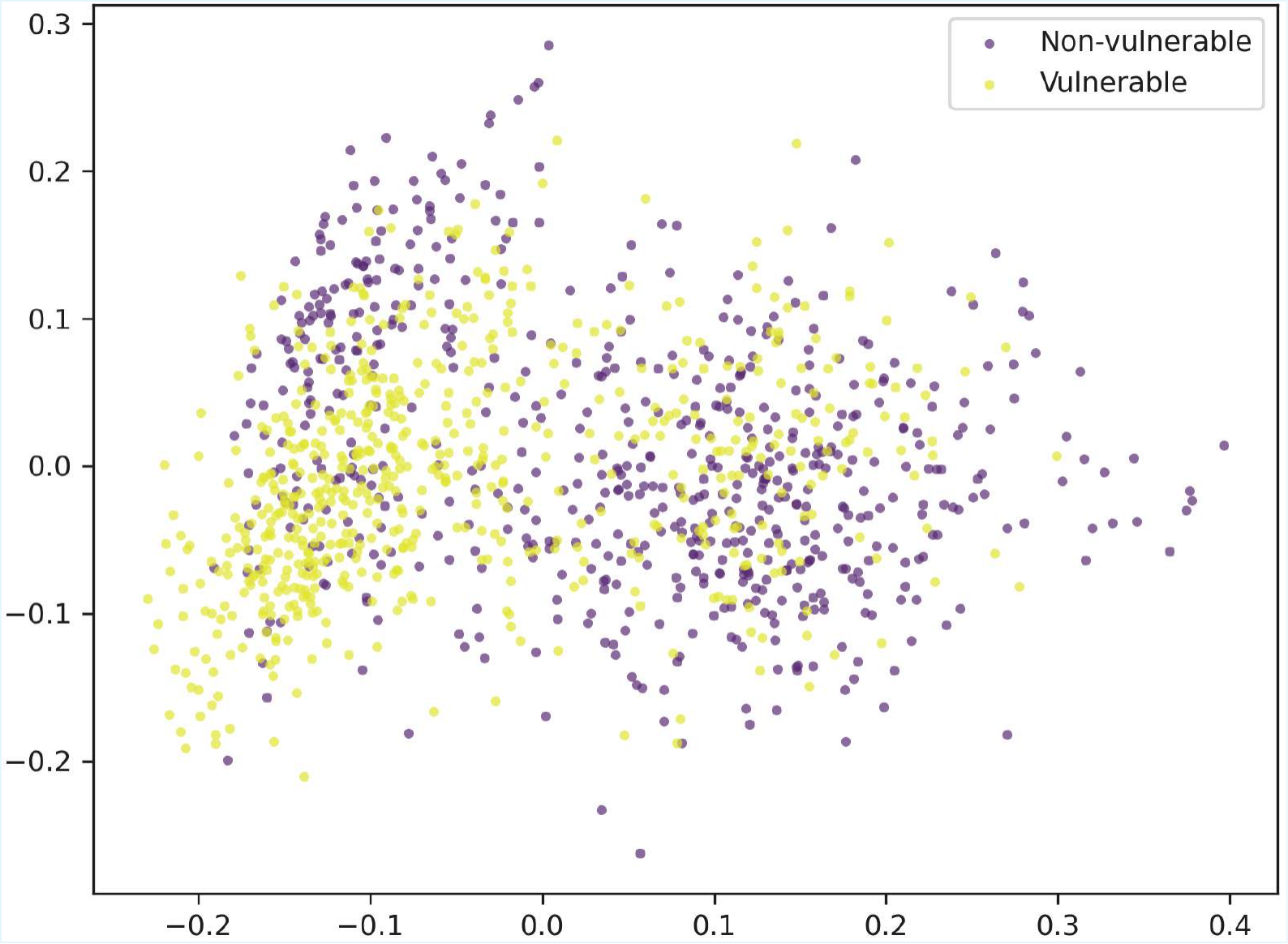}
\end{minipage}%
}%
\subfigure[Fine-Tuning]{
\begin{minipage}[t]{0.33\linewidth}
\centering
\includegraphics[width=1.0\linewidth]{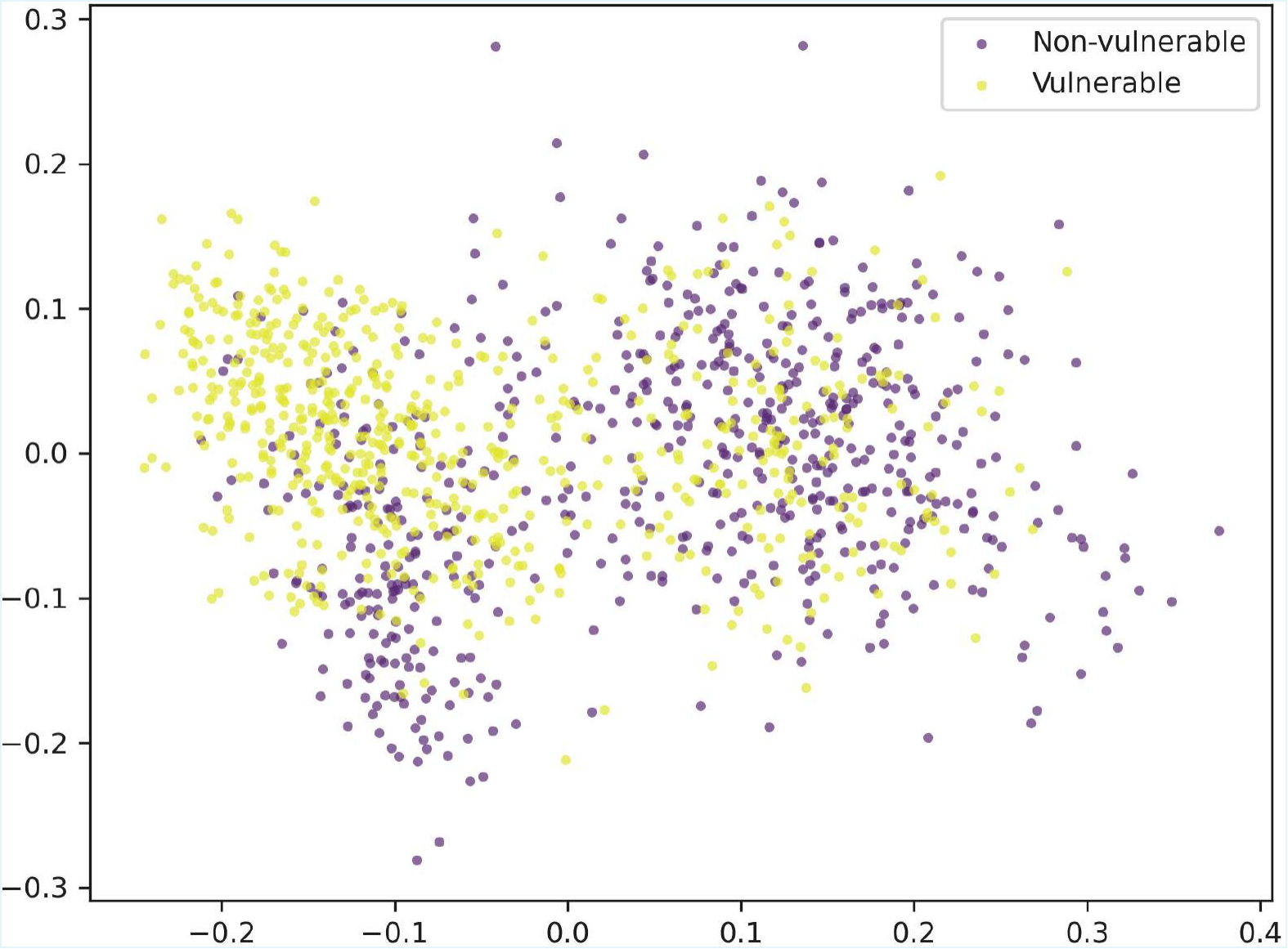}
\end{minipage}%
}%
\centering
\subfigure[MultiVul]{
\begin{minipage}[t]{0.33\linewidth}
\centering
\includegraphics[width=1.0\linewidth]{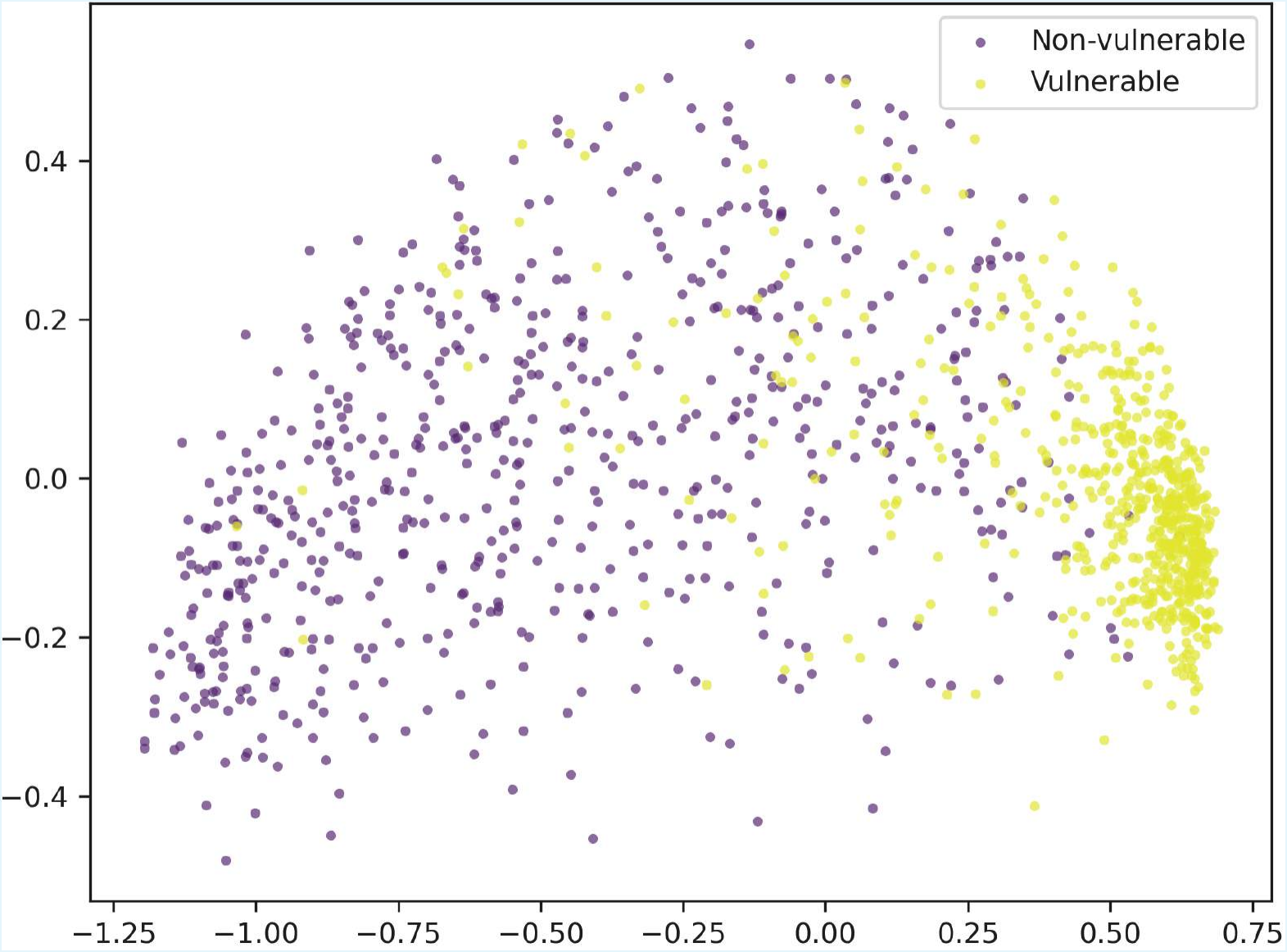}
\end{minipage}%
}%
\\
\centering
\subfigure[Zero-shot]{
\begin{minipage}[t]{0.33\linewidth}
\centering
\includegraphics[width=1.0\linewidth]{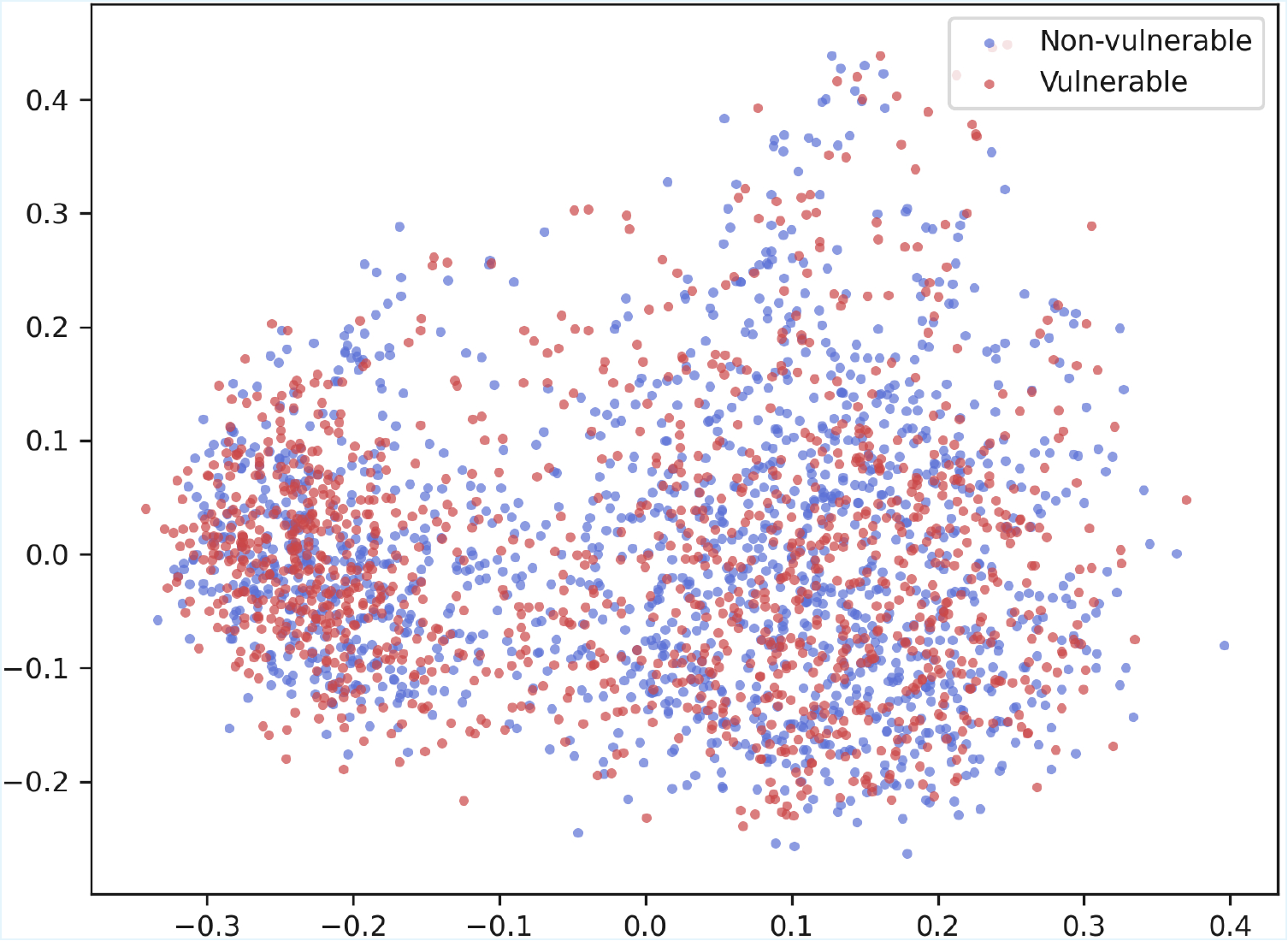}
\end{minipage}%
}%
\centering
\subfigure[Fine-Tuning]{
\begin{minipage}[t]{0.33\linewidth}
\centering
\includegraphics[width=1.0\linewidth]{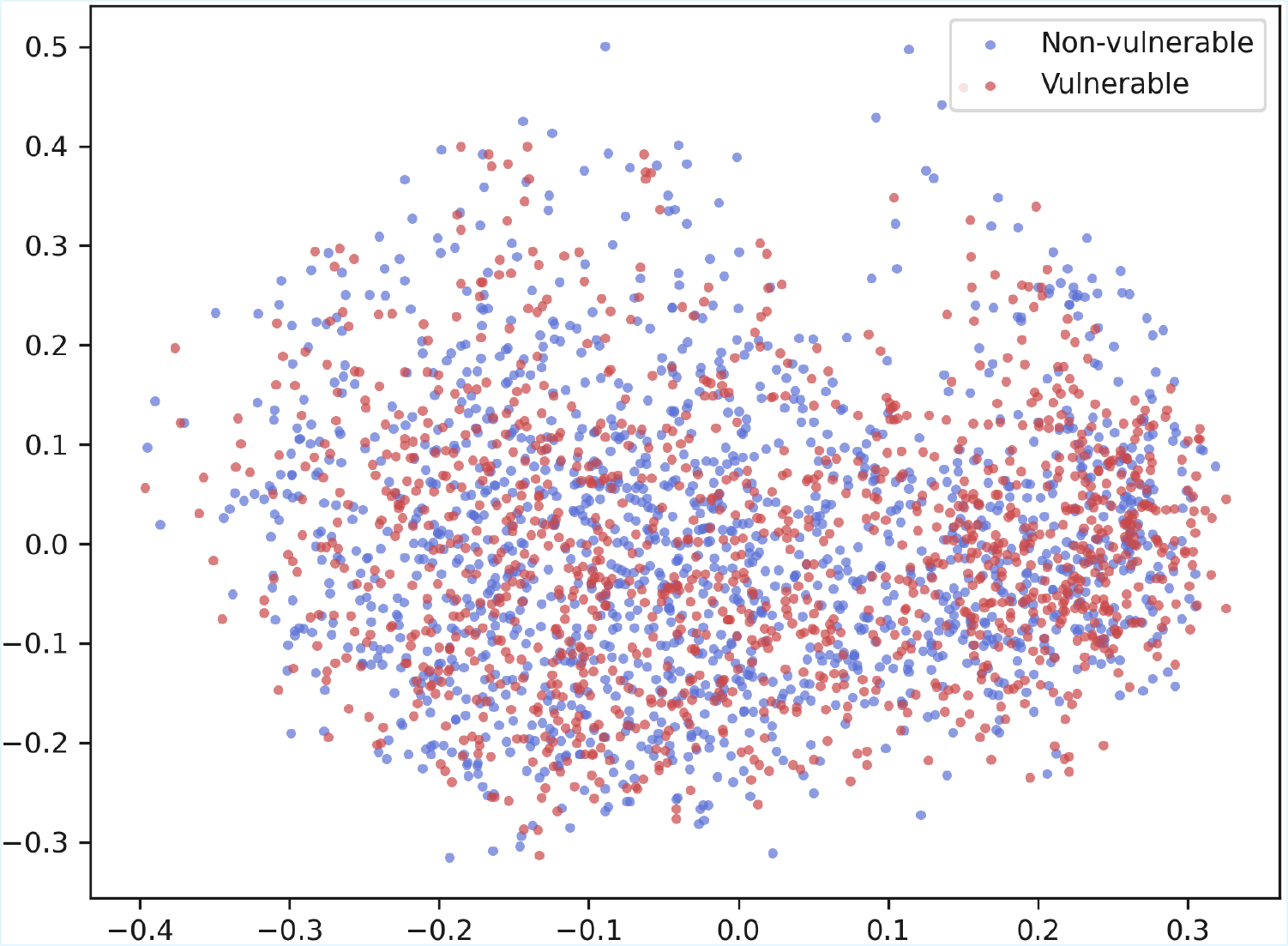}
\end{minipage}%
}%
\centering
\subfigure[MultiVul]{
\begin{minipage}[t]{0.33\linewidth}
\centering
\includegraphics[width=1.0\linewidth]{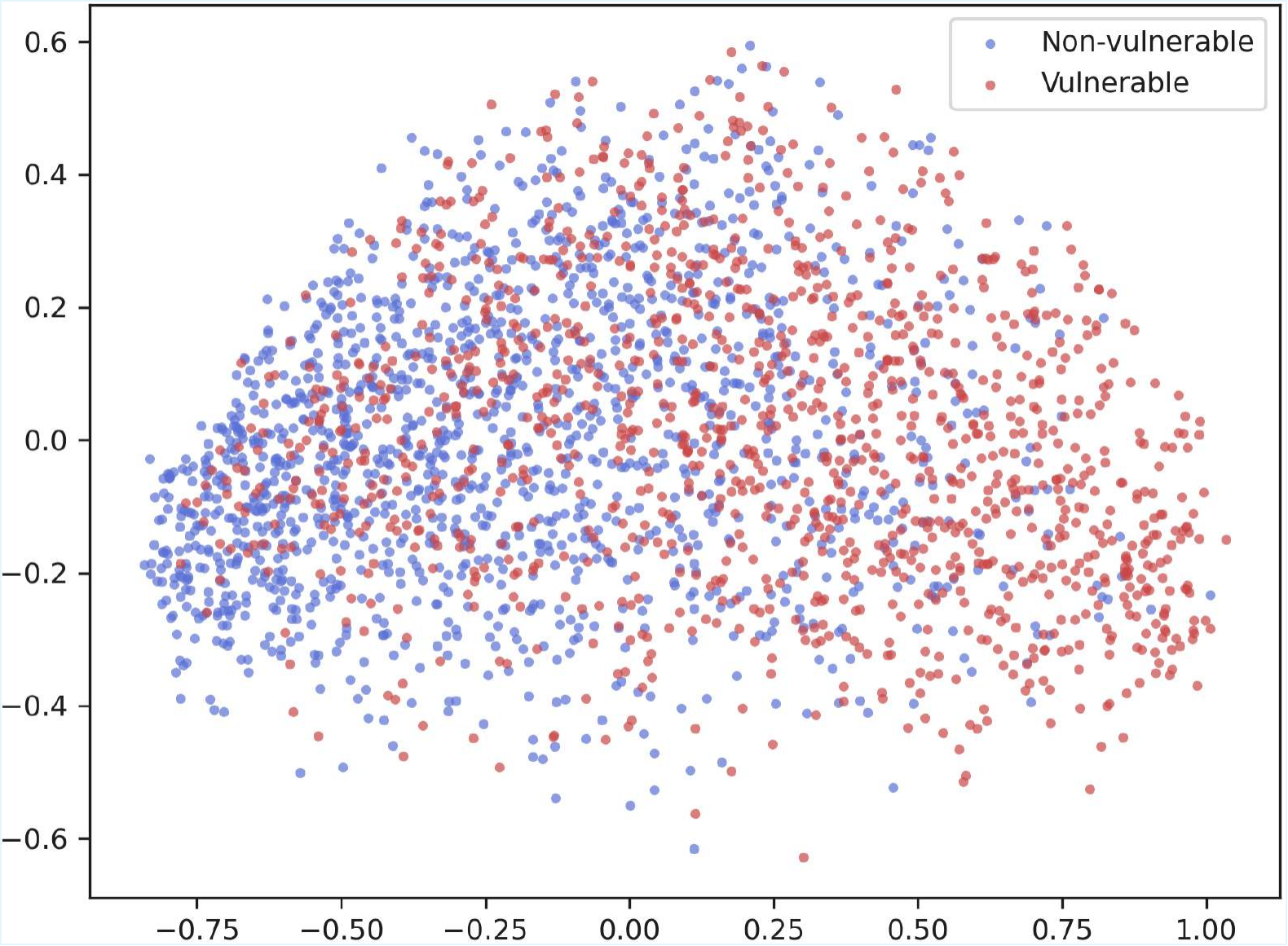}
\end{minipage}%
}%
\caption{Visualization of code embeddings after dimension reduction using Principal Component Analysis (PCA). Figures (a)–(c) show embeddings from Zero-shot, Fine-Tuning, and \textsc{MultiVul} on DiverseVul, while Figures (d)–(f) present the corresponding results on Devign. Model: Qwen2.5-Coder-7B.
}
\label{fig:PCA}
\end{figure*}

\textbf{Visualizing principal component analysis.} To further understand why \textsc{MultiVul} achieves superior performance, we visualize the code embeddings after dimensionality reduction on the testing dataset of DiverseVul and Devign using \emph{Principal Component Analysis (PCA)}, as shown in Figure~\ref{fig:PCA}.   Figures~(a)-(c) show the results on DiverseVul using Qwen2.5-Coder-7B, while Figures~(d)-(f) show the corresponding results on Devign using the same code LLM. In both datasets, the Zero-shot embeddings form highly overlapping clusters where vulnerable and non-vulnerable samples are mixed, indicating that the pretrained code LLM lacks task-specific discriminative capacity. Fine-tuning produces clearer separation but still exhibits substantial overlap along the decision boundary, suggesting that single-modal supervision (i.e., code only) learns partial semantic distinctions but fails to capture deeper vulnerability semantics. In contrast, \textsc{MultiVul} yields a more visually separable two-dimensional projection, where vulnerable and non-vulnerable samples exhibit less overlap than the baselines. This visualization suggests that the multimodal alignment (i.e., code and its textual comments) effectively enhances the semantic grounding of LLMs~\cite{cicchetti2025gramian}. By incorporating the paired textual descriptions, \textsc{MultiVul} captures semantic intent that is difficult to infer from code tokens alone, e.g., error handling in memory operations. 

Overall, \textsc{MultiVul} produces a more structured and separable embedding space, consistent with the Accuracy and F1 gains observed over both prompting-based and fine-tuning baselines.

\begin{answerbox}
\textit{\textbf{Answer to RQ1:} \textsc{MultiVul} consistently outperforms prompting-based and code-only \emph{Fine-Tuning} baselines across four code LLMs on DiverseVul and Devign, with up to 27.07\% and 13.37\% F1 improvements, respectively. The PCA results further show that \textsc{MultiVul} learns more separable vulnerability representations.}
\end{answerbox}

\subsection{RQ2: Ablation Study}
\label{sec:evaluation_ablation}
We further analyze the contributions of the two key components in \textsc{MultiVul}, namely ~\emph{Augmented Code--Text Alignment} and~\emph{Cross-View Consistency Regularization}, as illustrated in Figure~\ref{fig:overview}. Starting from standard Fine-Tuning, we first add the original code--text alignment (\emph{w/o Augmented Alignment}), then further introduce the augmented alignment while removing consistency regularization (\emph{w/o Consistency}), and finally obtain \textsc{MultiVul} by enabling both components. Overall, the full training method, i.e., \textsc{MultiVul}, delivers the strongest and most consistent performance across code LLMs and datasets.

Table~\ref{tab:ablation_id} shows that, on DiverseVul, \textsc{MultiVul} consistently outperforms both ablated variants across all four code LLMs in terms of both Accuracy and F1. This shows that augmented alignment and cross-view consistency provide complementary benefits. Compared with \emph{w/o Augmented Alignment}, \textsc{MultiVul} improves F1 by up to 3.06\%, and compared with \emph{w/o Consistency}, the gain reaches up to 3.91\%. A similar trend is observed for Accuracy, with gains reaching up to 2.40\% over \emph{w/o Augmented Alignment} and 3.52\% over \emph{w/o Consistency}. Overall, these results indicate that both augmented alignment and cross-view consistency contribute positively to DiverseVul, with the effect of cross-view consistency being especially evident on StarCoder2-7B, where removing consistency leads to a notable drop in both Accuracy and F1.

\begin{table*}[!ht]
\scriptsize
\setlength{\tabcolsep}{1.6pt}
\renewcommand{\arraystretch}{0.8}
\centering
\caption{Ablation study on DiverseVul and Devign across four code LLMs.
All metrics are reported in percentage (\%). \emph{Fine-Tuning} refers to standard supervised fine-tuning on code inputs only.
\emph{w/o Augmented Alignment} removes the augmented code--text alignment.
\emph{w/o Consistency} removes the cross-view consistency regularization.
Bold marks the best value per column within each code LLM. Blue shading highlights the highest Accuracy and F1 for each code LLM--dataset pair.}

\resizebox{\textwidth}{!}{
\begin{tabular}{lcccc|cccc|cccc|cccc}
\toprule
\multirow{3}{*}{\textbf{Methods}} &
\multicolumn{4}{c|}{\textbf{DeepSeek-Coder-6.7B}} &
\multicolumn{4}{c|}{\textbf{Qwen2.5-Coder-7B}} &
\multicolumn{4}{c|}{\textbf{StarCoder2-7B}} &
\multicolumn{4}{c}{\textbf{CodeLlama-7B}} \\
\cmidrule(lr){2-5}\cmidrule(lr){6-9}\cmidrule(lr){10-13}\cmidrule(lr){14-17}
& \multicolumn{2}{c}{\textbf{DiverseVul}} & \multicolumn{2}{c|}{\textbf{Devign}}
& \multicolumn{2}{c}{\textbf{DiverseVul}} & \multicolumn{2}{c|}{\textbf{Devign}}
& \multicolumn{2}{c}{\textbf{DiverseVul}} & \multicolumn{2}{c|}{\textbf{Devign}}
& \multicolumn{2}{c}{\textbf{DiverseVul}} & \multicolumn{2}{c}{\textbf{Devign}} \\
\cmidrule(lr){2-3}\cmidrule(lr){4-5}
\cmidrule(lr){6-7}\cmidrule(lr){8-9}
\cmidrule(lr){10-11}\cmidrule(lr){12-13}
\cmidrule(lr){14-15}\cmidrule(lr){16-17}
& Accuracy & F1 & Accuracy & F1
& Accuracy & F1 & Accuracy & F1
& Accuracy & F1 & Accuracy & F1
& Accuracy & F1 & Accuracy & F1 \\
\midrule
Fine-Tuning
& 78.72 & 77.61 & 56.19 & 29.64
& 88.72 & 88.18 & 63.07 & 61.88
& 84.00 & 83.25 & 62.20 & 51.87
& 87.28 & 86.91 & 60.51 & 48.62 \\

w/o Augmented Alignment
& 89.60 & 89.29 & 58.90 & 23.05
& 92.64 & 92.73 & 65.86 & 66.18
& 89.44 & 89.37 & 62.12 & 41.98
& 91.28 & 91.31 & 64.03 & 46.69 \\

w/o Consistency
& 89.84 & 89.55 & 58.10 & 18.05
& 92.96 & 92.70 & 65.20 & 65.95
& 88.32 & 88.52 & 62.64 & 55.73
& 91.52 & 91.52 & 63.74 & 45.78 \\

\cellcolor[HTML]{DBE7FC}\textbf{MultiVul}
& \cellcolor[HTML]{DBE7FC}\textbf{90.00} & \cellcolor[HTML]{DBE7FC}\textbf{90.98}
& \cellcolor[HTML]{DBE7FC}\textbf{60.95} & \cellcolor[HTML]{DBE7FC}\textbf{41.36}
& \cellcolor[HTML]{DBE7FC}\textbf{93.60} & \cellcolor[HTML]{DBE7FC}\textbf{93.59}
& \cellcolor[HTML]{DBE7FC}\textbf{67.11} & \cellcolor[HTML]{DBE7FC}\textbf{70.01}
& \cellcolor[HTML]{DBE7FC}\textbf{91.84} & \cellcolor[HTML]{DBE7FC}\textbf{92.43}
& \cellcolor[HTML]{DBE7FC}\textbf{64.69} & \cellcolor[HTML]{DBE7FC}\textbf{59.29}
& \cellcolor[HTML]{DBE7FC}\textbf{91.68} & \cellcolor[HTML]{DBE7FC}\textbf{91.54}
& \cellcolor[HTML]{DBE7FC}\textbf{64.30} & \cellcolor[HTML]{DBE7FC}\textbf{53.22} \\
\bottomrule
\end{tabular}
}
\label{tab:ablation_id}
\end{table*}

On Devign, the effects of \emph{w/o Augmented Alignment} and \emph{w/o Consistency} become even more pronounced. \textsc{MultiVul} again achieves the best Accuracy and F1 across all four code LLMs, and the gaps relative to the ablated variants are substantially larger than on DiverseVul. For instance, for DeepSeek-Coder-6.7B, F1 increases from 23.05\% with \emph{w/o Augmented Alignment} and 18.05\% with \emph{w/o Consistency} to 41.36\% with \textsc{MultiVul}, corresponding to gains of 18.31\% and 23.31\%, respectively. This indicates that neither partial variant is sufficient on this more challenging dataset. A similar phenomenon is observed for StarCoder2-7B, where F1 rises from 41.98\% and 55.73\% to 59.29\%. Qwen2.5-Coder-7B is comparatively more stable, but \textsc{MultiVul} still achieves the best overall result, improving F1 from 66.18\% and 65.95\% to 70.01\%. The Accuracy trends are consistent with the F1 results, with gains reaching up to 2.57\% over \emph{w/o Augmented Alignment} and 2.85\% over \emph{w/o Consistency}.

In conclusion, these ablation results show that augmented code--text alignment and cross-view consistency are both important components of \textsc{MultiVul}. Augmented alignment enriches the supervision signal by introducing stronger multi-view correspondence between code and text, while consistency regularization stabilizes learning across views. Their combination produces the most reliable improvements in both Accuracy and F1, especially on the more challenging Devign dataset, where partial multimodal designs are considerably less stable. Additionally, to keep the main text focused, we report only Accuracy and F1 in Table~\ref{tab:ablation_id} as they most directly summarize overall effectiveness and the balance between Precision and Recall. The complete results, including Precision and Recall, are provided in Appendix~\ref{appendix:detailed_resutlts_ablation}.

\begin{answerbox}
\textit{\textbf{Answer to RQ2:} Both augmented code--text alignment and cross-view consistency contribute to \textsc{MultiVul}. For instance, on Devign, the largest F1 gains reach 18.31\% over removing augmented alignment and 23.31\% over removing consistency.}
\end{answerbox}

\subsection{RQ3: Out-of-Distribution Generalization}
\label{sec:evaluation_ood}

Different from \emph{in-distribution (ID)} evaluation in Section~\ref{sec:evaluation_ablation}, where the training, validation, and testing datasets are drawn from the same benchmark split, \emph{out-of-distribution (OOD)} evaluation in our study focuses on cross-dataset generalization. Specifically, we train on Devign and evaluate on the DiverseVul testing dataset, and vice versa. We adopt this protocol because cross-project and cross-dataset transfer provides a realistic and widely used setting for assessing the performance of OOD generalization in software vulnerability detection~\cite{nguyen2024deep,zhang2023cpvd,du2024generalization}, where distribution shifts naturally arise from differences in coding standards, library usage, and project structure. This design is particularly meaningful because Devign is constructed from four diversified C projects~\cite{zhou2019devign}, whereas DiverseVul spans a much broader range of projects and CWEs, thereby inducing a stronger cross-dataset shift~\cite{chen2023diversevul}.

Table~\ref{tab:ablation_ood} reports the OOD results under this cross-dataset protocol. The column labels indicate the source training dataset. For example, \emph{DiverseVul} means training on DiverseVul and testing on Devign. Overall, \textsc{MultiVul} consistently achieves the best performance in terms of Accuracy and F1 across all four code LLMs under both cross-dataset settings, showing that the proposed multimodal training strategy improves generalization under distribution shift. When training on DiverseVul and testing on Devign, the gains over Fine-Tuning are moderate but consistent. The largest F1 gain appears with StarCoder2-7B, where F1 improves from 59.88\% to 63.40\%, yielding a 3.52\% increase. For Accuracy, CodeLlama-7B shows the largest gain, increasing from 47.55\% to 49.95\%, with a 2.40\% improvement. When training on Devign and testing on DiverseVul, the gains are generally larger. DeepSeek-Coder-6.7B obtains the largest F1 improvement, increasing from 33.02\% to 50.74\%, with a 17.72\% gain. CodeLlama-7B achieves the largest Accuracy gain, improving from 54.16\% to 59.92\%, producing a 5.76\% increase. These results suggest that \textsc{MultiVul} provides especially strong benefits under more challenging cross-dataset transfer.

\begin{table*}[!ht]
\scriptsize
\setlength{\tabcolsep}{1.6pt}
\renewcommand{\arraystretch}{0.8}
\centering
\caption{Ablation study of \textsc{MultiVul} for OOD detection across four code LLMs. All metrics are reported in percentage (\%). \emph{Fine-Tuning} refers to standard supervised fine-tuning on code inputs only.
\emph{w/o Augmented Alignment} removes the augmented code--text alignment.
\emph{w/o Consistency} removes the cross-view consistency regularization.
Bold marks the best value per column within each code LLM. Blue shading highlights the highest Accuracy and F1 for each code LLM--dataset pair.}

\resizebox{\textwidth}{!}{
\begin{tabular}{lcccc|cccc|cccc|cccc}
\toprule
\multirow{3}{*}{\textbf{Methods}} &
\multicolumn{4}{c|}{\textbf{DeepSeek-Coder-6.7B}} &
\multicolumn{4}{c|}{\textbf{Qwen2.5-Coder-7B}} &
\multicolumn{4}{c|}{\textbf{StarCoder2-7B}} &
\multicolumn{4}{c}{\textbf{CodeLlama-7B}} \\
\cmidrule(lr){2-5}\cmidrule(lr){6-9}\cmidrule(lr){10-13}\cmidrule(lr){14-17}
& \multicolumn{2}{c}{\textbf{DiverseVul}} & \multicolumn{2}{c|}{\textbf{Devign}}
& \multicolumn{2}{c}{\textbf{DiverseVul}} & \multicolumn{2}{c|}{\textbf{Devign}}
& \multicolumn{2}{c}{\textbf{DiverseVul}} & \multicolumn{2}{c|}{\textbf{Devign}}
& \multicolumn{2}{c}{\textbf{DiverseVul}} & \multicolumn{2}{c}{\textbf{Devign}} \\
\cmidrule(lr){2-3}\cmidrule(lr){4-5}
\cmidrule(lr){6-7}\cmidrule(lr){8-9}
\cmidrule(lr){10-11}\cmidrule(lr){12-13}
\cmidrule(lr){14-15}\cmidrule(lr){16-17}
& Accuracy & F1 & Accuracy & F1
& Accuracy & F1 & Accuracy & F1
& Accuracy & F1 & Accuracy & F1
& Accuracy & F1 & Accuracy & F1 \\
\midrule
Fine-Tuning
& 46.89 & 61.75 & 54.24 & 33.02
& 48.35 & 61.62 & 48.08 & 62.02
& 46.89 & 59.88 & 55.12 & 65.65
& 47.55 & 61.34 & 54.16 & 57.96 \\

w/o Augmented Alignment
& 46.15 & 61.66 & 53.12 & 37.66
& 48.28 & 62.21 & 49.92 & 65.87
& 47.03 & 60.69 & 55.84 & 63.97
& 49.60 & 62.11 & 53.60 & 62.53 \\

w/o Consistency
& 46.23 & 61.85 & 51.52 & 44.59
& 47.11 & 62.55 & 49.84 & 65.87
& 46.81 & 61.63 & 57.76 & 62.18
& 49.16 & 62.20 & 54.80 & 63.19 \\

\cellcolor[HTML]{DBE7FC}\textbf{MultiVul}
& \cellcolor[HTML]{DBE7FC}\textbf{47.25} & \cellcolor[HTML]{DBE7FC}\textbf{62.80}
& \cellcolor[HTML]{DBE7FC}\textbf{55.68} & \cellcolor[HTML]{DBE7FC}\textbf{50.74}
& \cellcolor[HTML]{DBE7FC}\textbf{49.13} & \cellcolor[HTML]{DBE7FC}\textbf{63.67}
& \cellcolor[HTML]{DBE7FC}\textbf{50.64} & \cellcolor[HTML]{DBE7FC}\textbf{66.29}
& \cellcolor[HTML]{DBE7FC}\textbf{47.45} & \cellcolor[HTML]{DBE7FC}\textbf{63.40}
& \cellcolor[HTML]{DBE7FC}\textbf{58.04} & \cellcolor[HTML]{DBE7FC}\textbf{67.87}
& \cellcolor[HTML]{DBE7FC}\textbf{49.95} & \cellcolor[HTML]{DBE7FC}\textbf{63.17}
& \cellcolor[HTML]{DBE7FC}\textbf{59.92} & \cellcolor[HTML]{DBE7FC}\textbf{63.77} \\
\bottomrule
\end{tabular}
}
\label{tab:ablation_ood}
\end{table*}

The ablation results further show that the two components play complementary roles under OOD shift. Specifically, removing~\emph{augmented alignment} generally weakens performance and makes the gains less stable. For example, when training on Devign and testing on DiverseVul, F1 on DeepSeek-Coder-6.7B decreases from 50.74\% to 37.66\%, a drop of 13.08\%. Accuracy on CodeLlama-7B also falls from 59.92\% to 53.60\%, corresponding to a drop of 6.32\%. These results suggest that the additional augmented code--text view provides useful supervision for learning more transferable representations.

Removing \emph{consistency} is often even more detrimental, especially in the more challenging transfer direction from Devign to DiverseVul. For example, F1 on DeepSeek-Coder-6.7B drops from 50.74\% to 44.59\%, decreasing by 6.15\%. Similarly, F1 on StarCoder2-7B declines from 67.87\% to 62.18\%, a decrease of 5.69\%. The largest drops in Accuracy follow the same trend, where Accuracy decreases from 59.92\% to 54.80\% on CodeLlama-7B, a drop of 5.12\%. These results suggest that cross-view consistency plays an important role in stabilizing multimodal learning under distribution shift.

Furthermore, we observe an interesting phenomenon. In several cases, training on DiverseVul and testing on Devign yields higher F1 than training and testing on Devign itself. For example, under \textsc{MultiVul}, the F1 score of DeepSeek-Coder-6.7B increases from 41.36\% in the ID setting to 62.80\% under cross-dataset transfer, while StarCoder2-7B improves from 59.29\% to 63.40\%, and CodeLlama-7B from 53.22\% to 63.17\%. This is plausible because DiverseVul provides broader project coverage and richer vulnerability semantics, which may help the model learn more transferable representations than those obtained from the narrower training distribution of Devign~\cite{herbold2018comparative}.

In summary, \textsc{MultiVul} improves not only ID effectiveness but also OOD generalization. By combining augmented code--text alignment with cross-view consistency, it yields the most reliable performance across diverse code LLMs.  Detailed results are reported in Appendix~\ref{appendix:detailed_resutlts_ood}.

\begin{answerbox}
\textit{\textbf{Answer to RQ3:} \textsc{MultiVul} consistently improves cross-dataset generalization across four code LLMs.  The ablation results further show that augmented alignment and consistency regularization are both important for stable performance under distribution shift.}
\end{answerbox}

\subsection{RQ4: Sensitivity Analysis}
\label{sec:evaluation_sensitivity}

For sensitivity analysis, we use Qwen2.5-Coder on DiverseVul as a representative configuration. Qwen2.5-Coder is one of the strongest and most stable code LLMs in the main results shown from Section~\ref{sec:evaluation_effectiveness}, while DiverseVul serves as a primary benchmark and exhibits clearer and more consistent performance trends. We analyze the sensitivity of \textsc{MultiVul} to two important factors, including the scale of the code LLM used for the task of code comment generation and the data augmentation strength $\alpha$. 

\begin{figure}[h]
    \centering

    \begin{minipage}[t]{0.49\columnwidth}
        \centering
        \includegraphics[width=\linewidth]{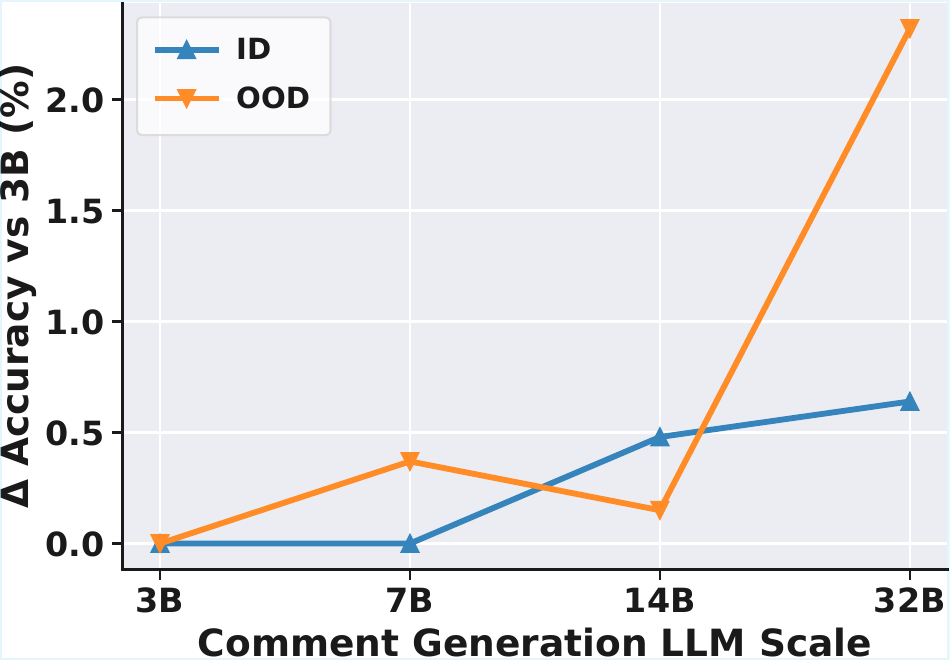}
        (a) $\Delta$ Accuracy vs 3B (\%).
        \label{fig:scale_acc}
    \end{minipage}
    \hfill
    \begin{minipage}[t]{0.49\columnwidth}
        \centering
        \includegraphics[width=\linewidth]{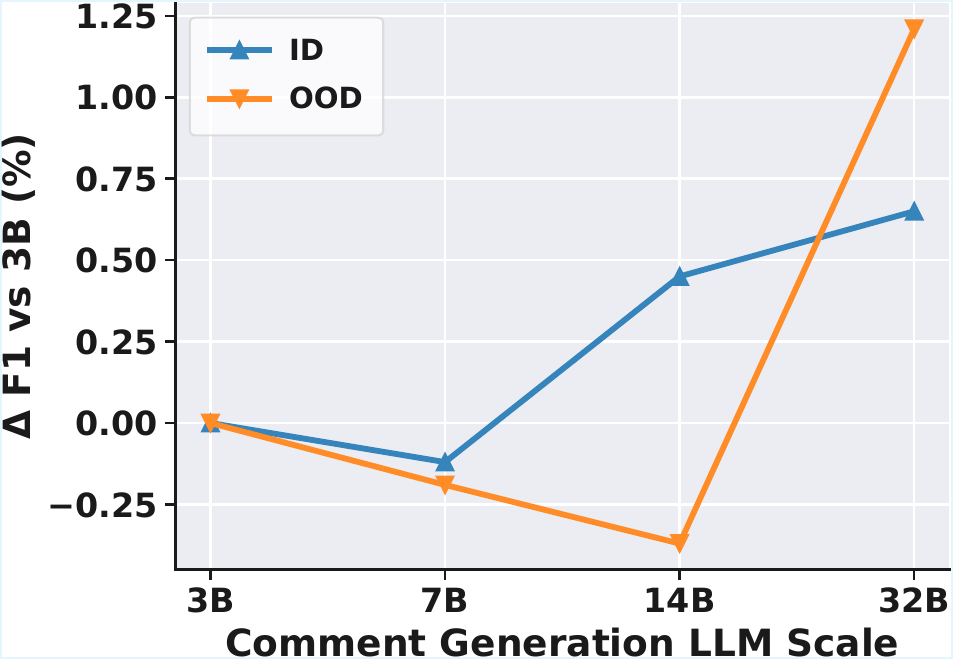}
        (b) $\Delta$ F1 vs 3B (\%).
        \label{fig:scale_f1}
    \end{minipage}

    \vspace{1mm}

    \begin{minipage}[t]{0.49\columnwidth}
        \centering
        \includegraphics[width=\linewidth]{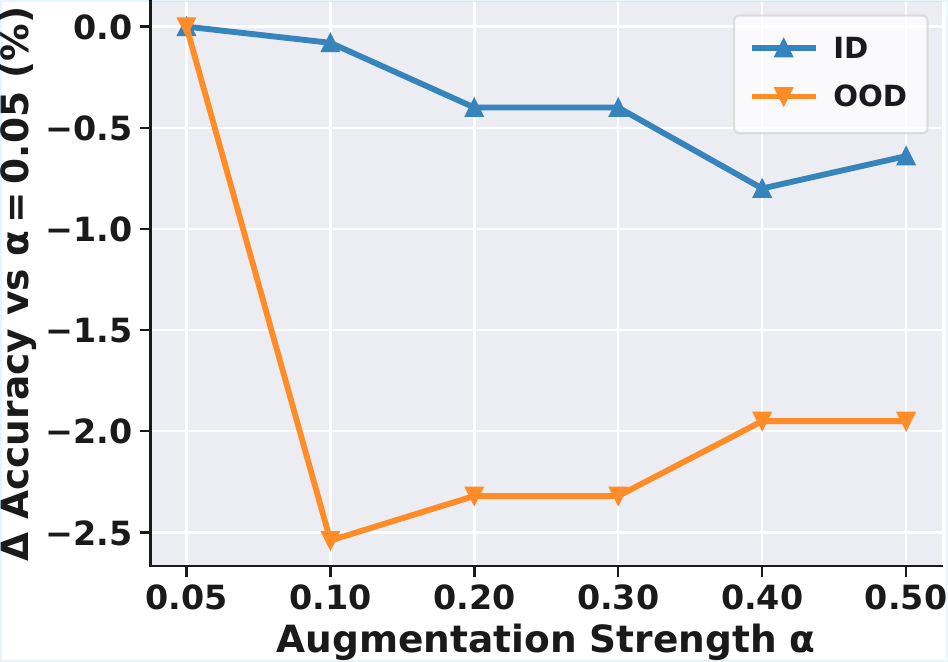}
        (c)$\Delta$ Accuracy vs $\alpha$=0.05.
        \label{fig:alpha_acc}
    \end{minipage}
    \hfill
    \begin{minipage}[t]{0.49\columnwidth}
        \centering
        \includegraphics[width=\linewidth]{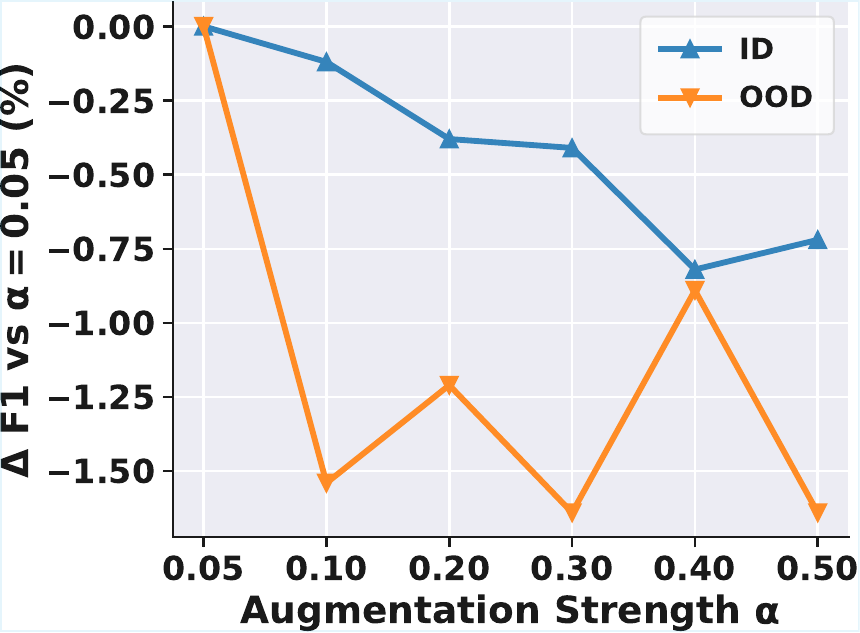}
         (d) $\Delta$ F1 vs $\alpha$=0.05.
        \label{fig:alpha_f1}
    \end{minipage}

    \caption{Sensitivity analysis of \textsc{MultiVul} on Qwen2.5-Coder over DiverseVul. The top row evaluates the effect on the scale of the comment generation LLM, while the bottom row evaluates the effect of data augmentation strength $\alpha$.}
    \label{fig:sensitivity}
\end{figure}

The top row of Figure~\ref {fig:sensitivity} shows that larger code LLMs used for code comment generation generally lead to better performance for \textsc{MultiVul}. On the~\emph{Effectiveness (ID detection)}, the gain is relatively modest. Compared with Qwen2.5-Coder-3B, Qwen2.5-Coder-32B improves Accuracy and F1 by only 0.64\% and 0.65\%, respectively. This suggests that \textsc{MultiVul} does not rely heavily on the size of code LLMs, as smaller models already provide competitive supervision. In contrast, the improvement is more pronounced on the~\emph{Generalization (OOD detection)}. Using Qwen2.5-Coder-32B increases Accuracy and F1 by 2.32\% and 1.21\% respectively, over Qwen2.5-Coder-3B. This indicates that higher-quality generated code comments are particularly beneficial for generalization under distribution shift.

As shown in the bottom row of Figure~\ref{fig:sensitivity}, \textsc{MultiVul} performs best under mild data augmentation, with $\alpha=0.05$ achieving the strongest overall results on both the \emph{Effectiveness (ID detection)} and \emph{Generalization (OOD detection)}. As $\alpha$ increases, performance generally declines, suggesting that stronger perturbations produced by data augmentation introduce noise and weaken vulnerability-relevant code--text alignment. This trend is especially clear in the OOD detection. For example, compared with $\alpha=0.05$, setting $\alpha=0.30$ reduces the accuracy and F1 of generalization by 2.32\% and 1.64\%, respectively, while $\alpha=0.50$ still leads to drops of 1.95 and 1.64\%. A similar but milder trend is observed in the ID detection. For instance, when $\alpha$ increases from 0.05 to 0.40, the accuracy and F1 of ID detection decrease by 0.80\% and 0.82\%, respectively. These results indicate that data augmentation provides useful regularization, whereas overly strong perturbations may distort the original semantics and reduce both effectiveness and generalization~\cite{rebuffi2021data,dong2023mixcode,dong2024effectiveness}.

In summary, these results show that \textsc{MultiVul} remains relatively stable across different scales of code LLMs, while being more sensitive to the choice of data augmentation strength. In particular, mild perturbation yields the best overall performance, whereas stronger data augmentation tends to introduce noise and weaken both effectiveness and generalization. For completeness, the full ID and OOD results are provided in Appendix~\ref{appendix:detailed_resutlts_sens}.

\begin{answerbox}
\textit{\textbf{Answer to RQ4:} Compared with 3B, using 32B improves ID Accuracy by only 0.64\%, but improves OOD Accuracy by 2.32\%, suggesting that code comments generated from larger-scale  LLMs are more useful under distribution shift. For data augmentation strength, $\alpha=0.05$ performs best, while stronger perturbations reduce both effectiveness and generalization.}
\end{answerbox}

\subsection{RQ5: Inference Latency}
\label{sec:evaluation_latency}

\begin{figure}[!h]
    \centering
\includegraphics[width=\columnwidth]{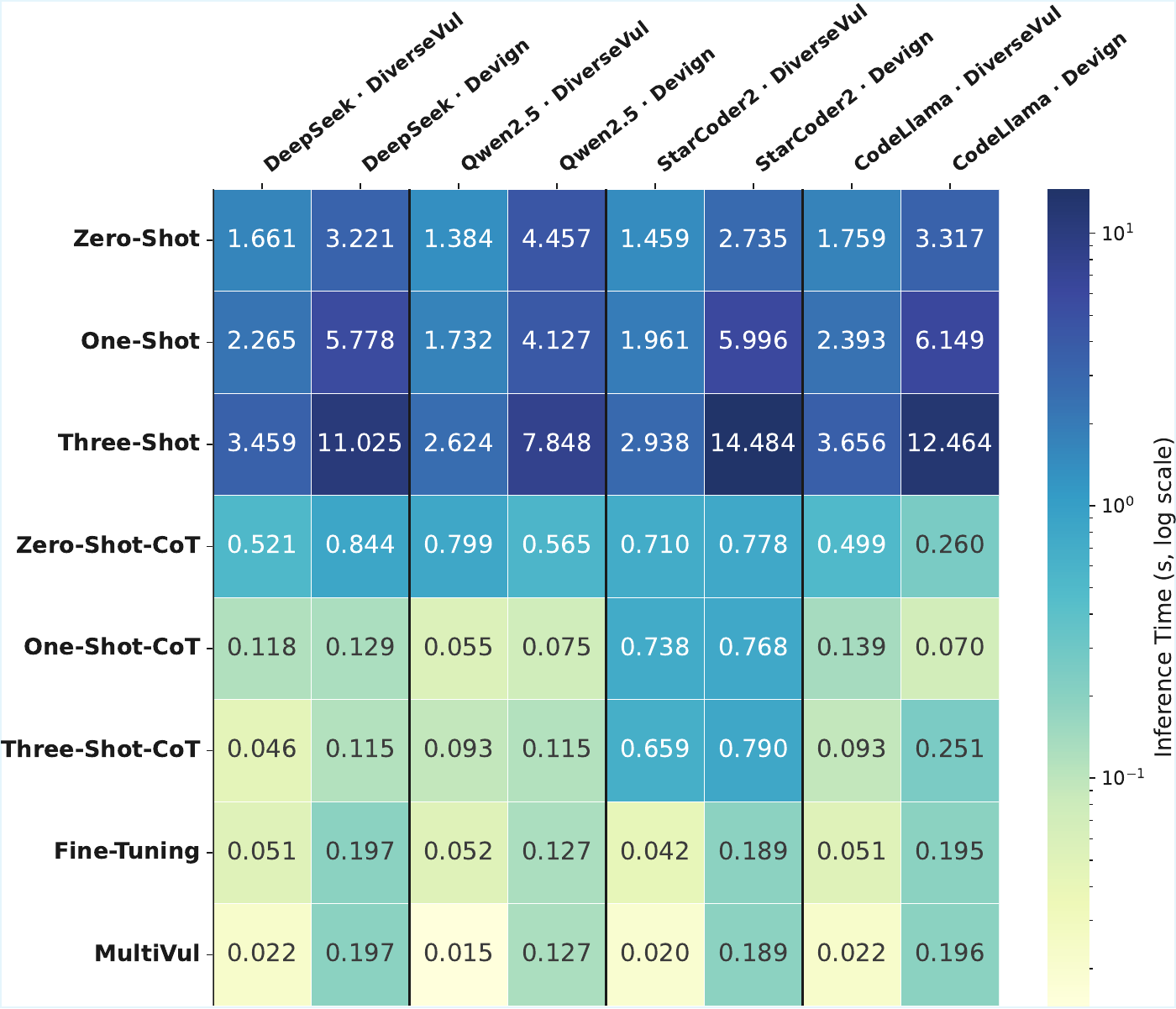}
    \caption{Inference latency across different methods.}
    \label{fig:inference}
\end{figure}

Figure~\ref{fig:inference} presents the inference latency of different methods across four code LLMs~\footnote{For simplicity, we refer to these models as \textit{DeepSeek}, \textit{Qwen2.5}, \textit{StarCoder2}, and \textit{CodeLlama}.} and two datasets. We can observe a clear trend that training-based methods, including \emph{Fine-Tuning} and \textsc{MultiVul}, are consistently much faster and more stable than prompting-based baselines. Prompting latency increases substantially as the number of in-context examples grows, with the highest cost appearing in the three-shot prompting. This effect is especially shown on Devign, where standard zero-shot, one-shot, and three-shot prompting ranges from $2.735$ to $14.484$ seconds per sample, whereas \textsc{MultiVul} requires only $0.127$ to $0.197$ seconds.


Importantly, \textsc{MultiVul} keeps the same deployment as standard fine-tuning. Although \textsc{MultiVul} leverages multimodal fusion and alignment during the training phase, the inference phase is performed using code inputs only, with a single forward pass through the encoder and a lightweight classifier head. It therefore avoids the expensive autoregressive decoding required by prompting-based methods (e.g., three-shot), whose latency scales with prompt length and generated output length~\cite{wang-etal-2025-agrec}. As a result, \textsc{MultiVul} remains comparable to \emph{Fine-Tuning} in inference cost while achieving roughly one to two orders of magnitude lower inference latency than standard prompting-based methods. This efficiency, together with its strong predictive performance, makes \textsc{MultiVul} more attractive for practical vulnerability detection scenarios where both effectiveness and deployment are important.
\begin{answerbox}
\textit{\textbf{Answer to RQ5:} \textsc{MultiVul} preserves efficient code-only inference. Its latency is comparable to \emph{Fine-Tuning}, and is roughly one to two orders of magnitude lower than prompting-based methods (e.g., one-shot), making it practical for deployment.}
\end{answerbox}

\subsection{Case Study}
\label{sec:case_study}
To better understand the behavior of \textsc{MultiVul} in vulnerability detection, we further conduct a qualitative analysis of false negative (FN) cases on DiverseVul using StarCoder2-7B.

\begin{table*}[!ht]
\centering
\caption{Error distribution by CWE category among false negatives (FN) on DiverseVul using StarCoder2-7B. \emph{CLIP} refers to the standard code--text alignment, without augmented code--text alignment or consistency regularization.}
\label{tab:cwe_fn_analysis}
\resizebox{2\columnwidth}{!}{
\begin{tabular}{lcccl}
\toprule
\textbf{CWE} & \textbf{Fine-Tuning} & \textbf{CLIP} & \textsc{MultiVul} & \textbf{Description} \\
\midrule
CWE-119 & 22 & 16 & 9 & Improper Restriction of Operations within the Bounds of a Memory Buffer \\
CWE-703 & 20 & 13 & 7 & Improper Check or Handling of Exceptional Conditions \\
CWE-20  & 19 & 14 & 9 & Improper Input Validation \\
CWE-416 & 18 & 8  & 5 & Use After Free \\
CWE-787 & 16 & 8  & 3 & Out-of-bounds Write \\
CWE-362 & 13 & 6  & 5 & Concurrent Execution using Shared Resource with Improper Synchronization \\
CWE-476 & 12 & 11 & 7 & NULL Pointer Dereference \\
CWE-200 & 10 & 8  & 6 & Exposure of Sensitive Information to an Unauthorized Actor \\
CWE-190 & 6  & 5  & 4 & Integer Overflow or Wraparound \\
CWE-617 & 4  & 4  & 0 & Reachable Assertion \\
Others  & 61 & 28 & 19 & Others \\
\bottomrule
\end{tabular}
}
\end{table*}

\begin{figure}[h]
    \centering
\includegraphics[width=0.65\columnwidth]{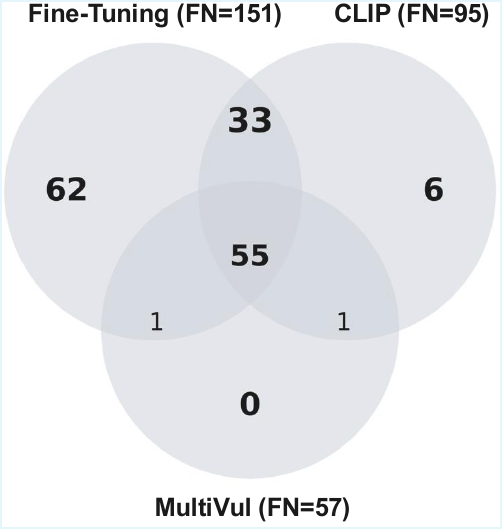}
  \caption{Venn diagram showing the overlap among the false-negative (FN) sets of \emph{Fine-Tuning}, \emph{CLIP}, and \textsc{MultiVul} on DiverseVul using StarCoder2-7B. \textsc{MultiVul} resolves 33 vulnerable functions that are missed by both \emph{Fine-Tuning} and \emph{CLIP}, while introducing no method-specific false negatives.}
\label{fig:case_study_venn}
\end{figure}

Figure~\ref{fig:case_study_venn} visualizes the overlap among the FN sets of Fine-Tuning, CLIP, and \textsc{MultiVul}. \emph{Fine-Tuning}, \emph{CLIP}, and \textsc{MultiVul} miss $151$, $95$, and $57$ vulnerable functions, respectively. Among the cases missed by both baselines, \textsc{MultiVul} correctly recovers $33$. In addition, there is no vulnerable function missed only by \textsc{MultiVul}. These results show that \textsc{MultiVul} reduces FNs by correctly identifying cases that both baselines miss, without introducing additional cases that only \textsc{MultiVul} fails to detect.

Table~\ref{tab:cwe_fn_analysis} reports the error distribution by CWE category among FNs. Overall, \textsc{MultiVul} shows clearer reductions on vulnerability categories that depend on contextual program behavior, such as exceptional-condition handling and memory-boundary reasoning. For example, on CWE-703, the number of missed cases decreases from $20$ under \emph{Fine-Tuning} and $13$ under \emph{CLIP} to $7$ under \textsc{MultiVul}. On CWE-119, they decrease from $22$ and $16$ to $9$. Similar reductions also appear for CWE-787 and CWE-416, involving memory-boundary and memory-lifetime errors, respectively.


These categories are difficult for code-only \emph{Fine-Tuning} because the vulnerability information is often distributed across validation checks, error-handling paths, buffer sizes, and memory operations, rather than being identifiable from a single local token\cite{zhou2019devign,fu2022linevul,cheng2022path}. \emph{CLIP} baseline improves over \emph{Fine-Tuning} by introducing code--text alignment, but it still aligns only the original code--text pair. Therefore, it lacks the key components, including augmented-view alignment and cross-view consistency, used in \textsc{MultiVul}. In contrast, \textsc{MultiVul} combines original code--text alignment, augmented code--text alignment, and consistency regularization. This design encourages the model to learn vulnerability-relevant semantics that remain stable across nearby code and text views, to help reduce the number of FNs in categories such as CWE-703 and CWE-119.

However, the improvement is not uniform across all CWE categories. For example, for CWE-190, FNs decrease only from $6$ and $5$ under \emph{Fine-Tuning} and \emph{CLIP} to $4$ under \textsc{MultiVul}. This suggests that integer-overflow vulnerabilities remain challenging for \textsc{MultiVul}. Unlike exceptional-condition or memory-boundary cases, CWE-190 often requires more precise numerical, range, or value-flow reasoning. The comments and lightweight data augmentations used by \textsc{MultiVul} can provide functional context and improve representation stability, but they do not explicitly model arithmetic constraints or value propagation. As a result, \textsc{MultiVul} shows more limited gains on this category than on categories that depend more on contextual program behavior.

\section{Discussion and Threats to Validity}
\label{sec:discussion}

\noindent
\textbf{Comment generation.}
\textsc{MultiVul} uses Qwen2.5-Coder-32B-Instruct to generate comments only during training, while prediction relies solely on source code. To reduce label leakage, the default prompt instructs the model not to mention security or vulnerabilities, and not to infer safety-related properties such as validation, error handling, permission checks, or safety guarantees unless explicitly present in the code. This code-only prediction design avoids requiring textual inputs during deployment and reduces the chance that comments expose vulnerability labels. Nevertheless, generated comments may still be incomplete or partially unsupported, which can affect code--text alignment. We mitigate this risk through critique prompting, where the model drafts a comment, checks unsupported claims, and produces a revised description grounded in the code.

\noindent
\textbf{Data augmentation.}
The RS and RD transformations used in \textsc{MultiVul} are lightweight perturbations and are not guaranteed to be strictly semantics-preserving for all source code~\cite{dong2025boosting}. However, \textsc{MultiVul} does not treat augmented code as an additional supervised training sample. Instead, augmented code and text are used to construct nearby views for contrastive alignment and consistency regularization. This design reduces the risk of directly learning from incorrectly labeled transformed programs and encourages more stable representations across local input variations.

\noindent
\textbf{Scope of comparisons.}
Our goal is to evaluate whether multimodal supervision improves the same code LLM under the same input setting. Therefore, our primary baselines use the same model, data splits, and code-only prediction setting. This controlled design isolates the effect of our training strategy from differences in model architecture or input information. Existing vulnerability detection methods, such as LineVul, VulBERTa, and CLeVeR, differ in model architecture, task formulation, model scale, or required inputs, making direct comparisons difficult to interpret. We therefore discuss them as related work rather than primary baselines.

\noindent
\textbf{Dataset and generalization limitations.}
We evaluate \textsc{MultiVul} on DiverseVul and Devign with four code LLMs, covering both within-dataset and cross-dataset evaluations, ablations, sensitivity analysis, and latency measurement. Cross-dataset evaluation provides a practical test of distribution shift, but it cannot cover all real-world deployment scenarios. Public vulnerability datasets may contain noisy labels, project-specific artifacts, or near-duplicate code fragments, which can affect the reported results. Broader evaluations across additional programming languages and repository-level detection remain valuable future directions.

\section{Conclusion}

This paper presented \textsc{MultiVul}, a multimodal vulnerability detection framework that uses automatically generated code comments as training-time supervision while preserving code-only inference. \textsc{MultiVul} aligns original and augmented code--text pairs using dual-CLIP losses and stabilizes the learned representation space through cross-view consistency regularization. Experiments on two benchmarks and four code LLMs show that \textsc{MultiVul} improves detection effectiveness and OOD generalization while maintaining efficient code-only inference, and ablations further confirm the importance of augmented alignment and consistency regularization.

These results indicate that multimodal supervision can strengthen vulnerability detection without requiring natural language input at inference time. Future work will explore richer training-time context, such as commit messages, issue reports, and vulnerability advisories, to further improve cross-project and cross-dataset generalization.

\bibliographystyle{ACM-Reference-Format}
\bibliography{reference}

\appendix

\section{Critique Prompting}
\label{sec:appendix_prompting}

For code comment generation, we use a critique prompting that improves factuality while avoiding unsupported or hallucinated descriptions. Specifically, we use a three-step critique prompting strategy to generate code comments in \textsc{MultiVul}. Given a function, the code LLM first produces a one-sentence summary of its core functionality. It is then prompted to critically review this draft and identify unsupported claims, missing core behavior, or statements not explicitly grounded in the code, etc. Finally, the code LLM revises the summary according to its self-critique and outputs a single sentence that describes only the behavior evidenced by the function.  The prompting process is defined as follows:

\begin{figure*}[h]
\centering
\begin{lstlisting}[style=arxivprompt]
System: "You are an expert code summarization assistant.

Hard constraints:
- Output exactly ONE sentence in English.
- Describe ONLY what is explicitly shown in the code.
- Do NOT claim input validation, error handling, bounds checks, permissions, or safety guarantees unless the code clearly shows them.
- Avoid speculative words such as 'ensure/ensures/ensuring', 'handle(s) errors', 'validate(s)', 'sanitize(s)', 'filter(s)', 'guarantee(s)' unless explicitly present.
- Do NOT mention security or vulnerabilities (this is the normal setting)."

User: "Please generate a short one-sentence comment describing the core functionality of the following function:
<code>
def calculate_factorial(n):
    if n == 0 or n == 1:
        return 1
    else:
        return n * calculate_factorial(n - 1)
</code>

Output ONLY the sentence."

Assistant (Draft): "Computes the factorial of n recursively, returning 1 when n is 0 or 1."

User: "Review your previous answer and list problems.

Check specifically for:
- Any speculation beyond the code (e.g., 'ensures', 'handles errors', 'validates', 'guarantees').
- Any claims of checks that are not explicitly shown (input validation, bounds checks, error handling, permissions).
- Missing core behavior (main operations, key calls, main data flow).

Output ONLY short bullet points. Do NOT revise yet."

Assistant (Review):
- No unsupported claims are made.
- The recursive call and base cases are clearly described.

User: "Based on the problems you found, improve your answer.

Requirements:
- Output exactly ONE sentence in English.
- Describe ONLY what is explicitly shown in the code.
- Remove any speculative or non-evidenced claims.
- Do NOT mention security or vulnerabilities.

Output ONLY the final sentence."

Assistant (Final): "Computes the factorial of n recursively with base cases for 0 and 1."
\end{lstlisting}
\caption{Code comment generation with critique prompting.}
\label{fig:appendix_prompt}
\end{figure*}

\section{Hyperparameters}
\label{sec:appendix_hyper}

The hyperparameters used for the \textsc{MultiVul} model training include various settings for model optimization and regularization, shown in Table~\ref{tab:hyper_Appendix}. 
\begin{table*}[ht]
\centering
\caption{Hyperparameters of \textsc{MultiVul}.}
\resizebox{\textwidth}{!}{
\begin{tabular}{lcp{10cm}}
\toprule
\textbf{Hyperparameter} & \textbf{Value} & \textbf{Description} \\
\midrule
Maximum Input Length & 4096 & Maximum number of tokens per input sequence. \\
Projection Dimension & 768 & Dimensionality of the shared space for code and text embeddings. \\
Batch Size & 8 & Number of samples per optimization step. \\
Training Epochs & 10 & Number of full passes over the training dataset. \\
Original-view CLIP Loss Weight & 0.5 & Weight of the contrastive loss for the original code--text view. \\
Augmented-view CLIP Loss Weight & 0.5 & Weight of the contrastive loss for the augmented code--text view. \\
Consistency Loss Weight & 0.1 & Weight of the consistency between the original view and the augmented view. \\
Learning Rate & 3e-05 & Optimizer step size. \\
Weight Decay & 0.0001 & Regularization via penalizing large weights. \\
\bottomrule
\end{tabular}
}
\label{tab:hyper_Appendix}
\end{table*}

\section{Detailed Results for Ablation Study}
\label{appendix:detailed_resutlts_ablation}

We provide the full results, including Accuracy, Precision, Recall, and F1, for the ablation experiments discussed in Section~\ref{sec:evaluation_ablation}. Specifically, Table~\ref{tab:appendix_ablation_id} reports the detailed results on the in-distribution (ID) testing, respectively. The table complements the main-text discussion by showing the complete performance of each ablated variant, including the effects of removing augmented-view alignment and consistency regularization.

\begin{table*}[h]
\scriptsize
\setlength{\tabcolsep}{1.8pt}
\renewcommand{\arraystretch}{0.8}
\centering
\caption{Ablation study on DiverseVul and Devign across four code LLMs.
All metrics are reported in percentage (\%). \emph{Fine-Tuning} refers to standard supervised fine-tuning on code inputs only.
\emph{w/o Augmented Alignment} removes the augmented code--text alignment.
\emph{w/o Consistency} removes the cross-view consistency regularization.
Bold marks the best value per column within each code LLM. Blue shading highlights the highest Accuracy and F1 for each code LLM--dataset pair.}

\resizebox{\textwidth}{!}{
\begin{tabular}{lcccccccc|cccccccc}
\toprule
\multirow{3}{*}{\textbf{Methods}} &
\multicolumn{8}{c|}{\textbf{DeepSeek-Coder-6.7B}} &
\multicolumn{8}{c}{\textbf{Qwen2.5-Coder-7B}} \\
\cmidrule(lr){2-9}\cmidrule(lr){10-17}
& \multicolumn{4}{c}{\textbf{DiverseVul}} & \multicolumn{4}{c|}{\textbf{Devign}}
& \multicolumn{4}{c}{\textbf{DiverseVul}} & \multicolumn{4}{c}{\textbf{Devign}} \\
\cmidrule(lr){2-5}\cmidrule(lr){6-9}\cmidrule(lr){10-13}\cmidrule(lr){14-17}
& Accuracy & Precision & Recall & F1 & Accuracy & Precision & Recall & F1
& Accuracy & Precision & Recall & F1 & Accuracy & Precision & Recall & F1 \\
\midrule
Fine-Tuning
& 78.72 & 81.88 & 73.76 & 77.61
& 56.19 & 55.51 & 20.22 & 29.64
& 88.72 & 92.61 & 84.16 & 88.18
& 63.08 & 58.51 & 65.65 & 61.88 \\
w/o Augmented Alignment
& 89.60 & 92.02 & 86.72 & 89.29
& 58.90 & 79.25 & 13.48 & 23.05
& 92.64 & 91.58 & \textbf{93.92} & 92.73
& 65.86 & 60.40 & 73.19 & 66.18 \\
w/o Consistency
& 89.84 & \textbf{92.20} & 87.04 & 89.55
& 58.10 & \textbf{84.00} & 10.11 & 18.05
& 92.96 & \textbf{96.21} & 89.44 & 92.70
& 65.20 & 59.59 & 73.84 & 65.95 \\
\cellcolor[HTML]{DBE7FC}\textbf{MultiVul}
& \cellcolor[HTML]{DBE7FC}\textbf{90.00} & \cellcolor[HTML]{DBE7FC}92.06 & \cellcolor[HTML]{DBE7FC}\textbf{89.92} & \cellcolor[HTML]{DBE7FC}\textbf{90.98}
& \cellcolor[HTML]{DBE7FC}\textbf{60.95} & \cellcolor[HTML]{DBE7FC}65.73 & \cellcolor[HTML]{DBE7FC}\textbf{30.18} & \cellcolor[HTML]{DBE7FC}\textbf{41.36}
& \cellcolor[HTML]{DBE7FC}\textbf{93.60} & \cellcolor[HTML]{DBE7FC}93.74 & \cellcolor[HTML]{DBE7FC}93.44 & \cellcolor[HTML]{DBE7FC}\textbf{93.59}
& \cellcolor[HTML]{DBE7FC}\textbf{67.11} & \cellcolor[HTML]{DBE7FC}\textbf{64.75} & \cellcolor[HTML]{DBE7FC}\textbf{76.21} & \cellcolor[HTML]{DBE7FC}\textbf{70.01} \\
\bottomrule
\end{tabular}
}

\resizebox{\textwidth}{!}{
\begin{tabular}{lcccccccc|cccccccc}
\toprule
\multirow{3}{*}{\textbf{Methods}} &
\multicolumn{8}{c|}{\textbf{StarCoder2-7B}} &
\multicolumn{8}{c}{\textbf{CodeLlama-7B}} \\
\cmidrule(lr){2-9}\cmidrule(lr){10-17}
& \multicolumn{4}{c}{\textbf{DiverseVul}} & \multicolumn{4}{c|}{\textbf{Devign}}
& \multicolumn{4}{c}{\textbf{DiverseVul}} & \multicolumn{4}{c}{\textbf{Devign}} \\
\cmidrule(lr){2-5}\cmidrule(lr){6-9}\cmidrule(lr){10-13}\cmidrule(lr){14-17}
& Accuracy & Precision & Recall & F1 & Accuracy & Precision & Recall & F1
& Accuracy & Precision & Recall & F1 & Accuracy & Precision & Recall & F1 \\
\midrule
Fine-Tuning
& 84.00 & 87.35 & 79.52 & 83.25
& 62.20 & 61.92 & 44.62 & 51.87
& 87.28 & 89.49 & 84.48 & 86.91
& 60.51 & 59.86 & 40.93 & 48.62 \\
w/o Augmented Alignment
& 89.44 & 89.95 & 88.80 & 89.37
& 62.12 & \textbf{69.78} & 30.02 & 41.98
& 91.28 & 90.95 & \textbf{91.68} & 91.31
& 64.03 & \textbf{72.15} & 34.51 & 46.69 \\
w/o Consistency
& 88.32 & 87.02 & 90.08 & 88.52
& 62.64 & 60.68 & 51.52 & 55.73
& 91.52 & 91.52 & 91.52 & 91.52
& 63.74 & 72.07 & 33.55 & 45.78 \\
\cellcolor[HTML]{DBE7FC}\textbf{MultiVul}
& \cellcolor[HTML]{DBE7FC}\textbf{91.84} & \cellcolor[HTML]{DBE7FC}\textbf{93.37} & \cellcolor[HTML]{DBE7FC}\textbf{91.52} & \cellcolor[HTML]{DBE7FC}\textbf{92.43}
& \cellcolor[HTML]{DBE7FC}\textbf{64.69} & \cellcolor[HTML]{DBE7FC}62.57 & \cellcolor[HTML]{DBE7FC}\textbf{56.34} & \cellcolor[HTML]{DBE7FC}\textbf{59.29}
& \cellcolor[HTML]{DBE7FC}\textbf{91.68} & \cellcolor[HTML]{DBE7FC}\textbf{93.06} & \cellcolor[HTML]{DBE7FC}90.08 & \cellcolor[HTML]{DBE7FC}\textbf{91.54}
& \cellcolor[HTML]{DBE7FC}\textbf{64.30} & \cellcolor[HTML]{DBE7FC}63.62 & \cellcolor[HTML]{DBE7FC}\textbf{45.75} & \cellcolor[HTML]{DBE7FC}\textbf{53.22} \\
\bottomrule
\end{tabular}
}
\label{tab:appendix_ablation_id}
\end{table*}

\section{Detailed Results for Out-of-Distribution Generalization} 
\label{appendix:detailed_resutlts_ood}

We further report the full results for the out-of-distribution (OOD) generalization experiments discussed in Section~\ref{sec:evaluation_ood}. These results from Table~\ref{tab:appendix_ablation_ood} provide a more detailed view of how different methods behave under cross-dataset evaluation, including Accuracy, Precision, Recall, and F1 on the OOD detection. They supplement the main-text analysis by showing the extent to which \textsc{MultiVul} maintains stronger generalization than prompting-based and code-only fine-tuning baselines under distribution shift.

\begin{table*}[h]
\scriptsize
\setlength{\tabcolsep}{1.8pt}
\renewcommand{\arraystretch}{0.8}
\centering
\caption{Ablation study of \textsc{MultiVul} for OOD detection across four code LLMs. All metrics are reported in percentage (\%). \emph{Fine-Tuning} refers to standard supervised fine-tuning on code inputs only.
\emph{w/o Augmented Alignment} removes the augmented code--text alignment.
\emph{w/o Consistency} removes the cross-view consistency regularization.
Bold marks the best value per column within each code LLM. Blue shading highlights the highest Accuracy and F1 for each code LLM--dataset pair.}

\resizebox{\textwidth}{!}{
\begin{tabular}{lcccccccc|cccccccc}
\toprule
\multirow{3}{*}{\textbf{Methods}} &
\multicolumn{8}{c|}{\textbf{DeepSeek-Coder-6.7B}} &
\multicolumn{8}{c}{\textbf{Qwen2.5-Coder-7B}} \\
\cmidrule(lr){2-9}\cmidrule(lr){10-17}
& \multicolumn{4}{c}{\textbf{DiverseVul}} & \multicolumn{4}{c|}{\textbf{Devign}}
& \multicolumn{4}{c}{\textbf{DiverseVul}} & \multicolumn{4}{c}{\textbf{Devign}} \\
\cmidrule(lr){2-5}\cmidrule(lr){6-9}\cmidrule(lr){10-13}\cmidrule(lr){14-17}
& Accuracy & Precision & Recall & F1 & Accuracy & Precision & Recall & F1
& Accuracy & Precision & Recall & F1 & Accuracy & Precision & Recall & F1 \\
\midrule
Fine-Tuning
& 46.89 & \textbf{46.15} & 93.26 & 61.75
& 54.24 & \textbf{61.57} & 22.56 & 33.02
& 48.35 & 46.62 & 90.85 & 61.62
& 48.08 & 48.89 & 84.80 & 62.02 \\
w/o Augmented Alignment
& 46.15 & 45.67 & 94.86 & 61.66
& 53.12 & 56.19 & 28.32 & 37.66
& 48.28 & 46.67 & 93.26 & 62.21
& 49.92 & 49.96 & 96.64 & 65.87 \\
w/o Consistency
& 46.23 & 45.73 & 95.51 & 61.85
& 51.52 & 51.69 & 39.20 & 44.59
& 47.11 & 46.21 & 96.79 & 62.55
& 49.84 & 49.92 & 96.80 & 65.87 \\
\cellcolor[HTML]{DBE7FC}\textbf{MultiVul}
& \cellcolor[HTML]{DBE7FC}\textbf{47.25} & \cellcolor[HTML]{DBE7FC}46.15 & \cellcolor[HTML]{DBE7FC}\textbf{98.23} & \cellcolor[HTML]{DBE7FC}\textbf{62.80}
& \cellcolor[HTML]{DBE7FC}\textbf{55.68} & \cellcolor[HTML]{DBE7FC}60.06 & \cellcolor[HTML]{DBE7FC}\textbf{43.92} & \cellcolor[HTML]{DBE7FC}\textbf{50.74}
& \cellcolor[HTML]{DBE7FC}\textbf{49.13} & \cellcolor[HTML]{DBE7FC}\textbf{47.17} & \cellcolor[HTML]{DBE7FC}\textbf{97.93} & \cellcolor[HTML]{DBE7FC}\textbf{63.67}
& \cellcolor[HTML]{DBE7FC}\textbf{50.64} & \cellcolor[HTML]{DBE7FC}\textbf{50.33} & \cellcolor[HTML]{DBE7FC}\textbf{97.04} & \cellcolor[HTML]{DBE7FC}\textbf{66.29} \\
\bottomrule
\end{tabular}
}

\resizebox{\textwidth}{!}{
\begin{tabular}{lcccccccc|cccccccc}
\toprule
\multirow{3}{*}{\textbf{Methods}} &
\multicolumn{8}{c|}{\textbf{StarCoder2-7B}} &
\multicolumn{8}{c}{\textbf{CodeLlama-7B}} \\
\cmidrule(lr){2-9}\cmidrule(lr){10-17}
& \multicolumn{4}{c}{\textbf{DiverseVul}} & \multicolumn{4}{c|}{\textbf{Devign}}
& \multicolumn{4}{c}{\textbf{DiverseVul}} & \multicolumn{4}{c}{\textbf{Devign}} \\
\cmidrule(lr){2-5}\cmidrule(lr){6-9}\cmidrule(lr){10-13}\cmidrule(lr){14-17}
& Accuracy & Precision & Recall & F1 & Accuracy & Precision & Recall & F1
& Accuracy & Precision & Recall & F1 & Accuracy & Precision & Recall & F1 \\
\midrule
Fine-Tuning
& 46.89 & 45.69 & 86.84 & 59.88
& 55.12 & 53.17 & \textbf{85.76} & 65.65
& 47.55 & 46.35 & 90.70 & 61.34
& 54.16 & 53.52 & 63.20 & 57.96 \\
w/o Augmented Alignment
& 47.03 & 45.89 & 89.57 & 60.69
& 55.84 & 54.02 & 78.40 & 63.97
& 49.60 & 47.28 & 90.53 & 62.11
& 53.60 & 52.44 & 77.44 & 62.53 \\
w/o Consistency
& 46.81 & 45.94 & 93.58 & 61.63
& 57.76 & 56.29 & 69.44 & 62.18
& 49.16 & 47.07 & 91.65 & 62.20
& 54.80 & 53.30 & 77.60 & 63.19 \\
\cellcolor[HTML]{DBE7FC}\textbf{MultiVul}
& \cellcolor[HTML]{DBE7FC}\textbf{47.45} & \cellcolor[HTML]{DBE7FC}\textbf{46.74} & \cellcolor[HTML]{DBE7FC}\textbf{98.52} & \cellcolor[HTML]{DBE7FC}\textbf{63.40}
& \cellcolor[HTML]{DBE7FC}\textbf{58.04} & \cellcolor[HTML]{DBE7FC}\textbf{57.31} & \cellcolor[HTML]{DBE7FC}83.21 & \cellcolor[HTML]{DBE7FC}\textbf{67.87}
& \cellcolor[HTML]{DBE7FC}\textbf{49.95} & \cellcolor[HTML]{DBE7FC}\textbf{47.59} & \cellcolor[HTML]{DBE7FC}\textbf{93.89} & \cellcolor[HTML]{DBE7FC}\textbf{63.17}
& \cellcolor[HTML]{DBE7FC}\textbf{59.92} & \cellcolor[HTML]{DBE7FC}\textbf{58.18} & \cellcolor[HTML]{DBE7FC}\textbf{70.56} & \cellcolor[HTML]{DBE7FC}\textbf{63.77} \\
\bottomrule
\end{tabular}
}
\label{tab:appendix_ablation_ood}
\end{table*}

\section{Detailed Results for Sensitivity Analysis}
\label{appendix:detailed_resutlts_sens}

We also provide the full results for the sensitivity analysis discussed in Section~\ref{sec:evaluation_sensitivity}. Specifically, Tables~\ref{tab:appendix_scale_qwen_diversevul} and~\ref{tab:appendix_alpha_qwen_diversevul} report the detailed results for varying the scale of the code comment generation LLM and the augmentation strength $\alpha$, respectively. These tables include Accuracy, Precision, Recall, and F1, and complement the main-text figures by showing the complete numerical trends underlying the sensitivity analysis.

\begin{table*}[h]
\scriptsize
\setlength{\tabcolsep}{4.0pt}
\renewcommand{\arraystretch}{1}
\centering
\caption{Sensitivity analysis of \textsc{MultiVul} to the scale of the code comment generation LLM on Qwen2.5-Coder  over DiverseVul. All metrics are reported in percentage (\%). Bold marks the best value in each column. Blue shading highlights the best Accuracy and F1 within each split.}
\resizebox{0.8\textwidth}{!}{
\begin{tabular}{lcccc|cccc}
\toprule
\multirow{2}{*}{\textbf{LLM Scale}} &
\multicolumn{4}{c|}{\textbf{ID (Effectiveness)}} &
\multicolumn{4}{c}{\textbf{OOD (Generalization)}} \\
\cmidrule(lr){2-5}\cmidrule(lr){6-9}
& Accuracy & Precision & Recall & F1 & Accuracy & Precision & Recall & F1 \\
\midrule
3B  & 92.96 & 93.24 & 92.64 & 92.94 & 46.81 & 46.07 & 96.95 & 62.46 \\
7B  & 92.96 & \textbf{94.68} & 91.04 & 92.82 & 47.18 & 46.20 & 95.51 & 62.27 \\
14B & 93.44 & 94.15 & 92.64 & 93.39 & 46.96 & 46.08 & 95.18 & 62.09 \\
\cellcolor[HTML]{DBE7FC}\textbf{32B}
    & \cellcolor[HTML]{DBE7FC}\textbf{93.60}
    & \cellcolor[HTML]{DBE7FC}93.74
    & \cellcolor[HTML]{DBE7FC}\textbf{93.44}
    & \cellcolor[HTML]{DBE7FC}\textbf{93.59}
    & \cellcolor[HTML]{DBE7FC}\textbf{49.13}
    & \cellcolor[HTML]{DBE7FC}\textbf{47.17}
    & \cellcolor[HTML]{DBE7FC}\textbf{97.93}
    & \cellcolor[HTML]{DBE7FC}\textbf{63.67} \\
\bottomrule
\end{tabular}
}
\label{tab:appendix_scale_qwen_diversevul}
\end{table*}

\begin{table*}[!t]
\scriptsize
\setlength{\tabcolsep}{4.0pt}
\renewcommand{\arraystretch}{1}
\centering
\caption{Sensitivity analysis of \textsc{MultiVul} to augmentation strength $\alpha$ on Qwen2.5-Coder over DiverseVul. All metrics are reported in percentage (\%). Bold marks the best value in each column. Blue shading highlights the best Accuracy and F1 within each split.}
\resizebox{0.8\textwidth}{!}{
\begin{tabular}{lcccc|cccc}
\toprule
\multirow{2}{*}{\textbf{Augmentation Strength $\alpha$}} &
\multicolumn{4}{c|}{\textbf{ID (Effectiveness)}} &
\multicolumn{4}{c}{\textbf{OOD (Generalization)}} \\
\cmidrule(lr){2-5}\cmidrule(lr){6-9}
& Accuracy & Precision & Recall & F1 & Accuracy & Precision & Recall & F1 \\
\midrule
\cellcolor[HTML]{DBE7FC}0.05
    & \cellcolor[HTML]{DBE7FC}\textbf{93.60}
    & \cellcolor[HTML]{DBE7FC}93.74
    & \cellcolor[HTML]{DBE7FC}\textbf{93.44}
    & \cellcolor[HTML]{DBE7FC}\textbf{93.59}
    & \cellcolor[HTML]{DBE7FC}\textbf{49.13}
    & \cellcolor[HTML]{DBE7FC}\textbf{47.17}
    & \cellcolor[HTML]{DBE7FC}\textbf{97.93}
    & \cellcolor[HTML]{DBE7FC}\textbf{63.67} \\
0.10 & 93.52 & \textbf{94.16} & 92.80 & 93.47 & 46.59 & 45.93 & 95.99 & 62.13 \\
0.20 & 93.20 & 93.13 & 93.28 & 93.21 & 46.81 & 46.07 & 96.95 & 62.46 \\
0.30 & 93.20 & 93.41 & 92.96 & 93.18 & 46.81 & 46.00 & 95.18 & 62.03 \\
0.40 & 92.80 & 93.21 & 92.32 & 92.77 & 47.18 & 46.27 & 97.59 & 62.78 \\
0.50 & 92.96 & 94.09 & 91.68 & 92.87 & 47.18 & 46.16 & 94.54 & 62.03 \\
\bottomrule
\end{tabular}
}
\label{tab:appendix_alpha_qwen_diversevul}
\end{table*}


\end{document}